\documentclass[prd,twocolumn,showpacs,amsmath,floatfix]{revtex4}
\usepackage{graphicx}
\begin{document}
\preprint{hep-ph/0210174}

\title{Radiative Corrections to Double Dalitz Decays:\\ 
       Effects on Invariant Mass Distributions and Angular Correlations}

\author{A. R. Barker}
\email[E--mail: ]{tonyb@cuhep.colorado.edu}
\author{H. Huang}
\author{P. A. Toale}
\author{J. Engle}
\altaffiliation[Current address: ]{Department of Physics,
                                   104 Davey Lab,
                                   Pennsylvania State University,
                                   University Park, PA 16802-6300.}
\affiliation{Department of Physics, University of Colorado\\
             Boulder, CO 80309-0390}

\date{\today}

\begin{abstract}
We review the theory of meson decays to two lepton pairs, including the
cases of identical as well as non--identical leptons, as well as
CP--conserving and CP--violating couplings. A complete lowest--order
calculation of QED radiative corrections to these decays is discussed, and
comparisons of predicted rates and kinematic distributions between
tree--level and one--loop--corrected calculations are presented for both
$\pz$ and $\kz$ decays.
\end{abstract}


\maketitle

\section{Background}
Meson decays to two photons should exhibit interesting correlations between
the photon polarizations\cite{yang}. Although existing particle detectors 
cannot measure photon polarizations directly, it has long been known that 
the polarization correlations can be measured indirectly by studying angular
correlations in the related double Dalitz decays\cite{kroll} in which both 
photons undergo internal conversion to a lepton pair. More recently, it has
been pointed out\cite{uy1} that a detailed study of these correlations can
be used to determine the relative amount of two possible meson--$\g\g$ 
couplings (one CP--conserving and one CP--violating for mesons that are CP 
eigenstates) that can contribute to this process.

Dalitz and double Dalitz decays are also of interest because they can be
exploited to perform a measurement of the electromagnetic form factor of the
decaying meson --- that is, how the meson couples to one real and one virtual
photon (Dalitz decay) or two virtual photons (double Dalitz) depends on the
$q^2$ values of the photon(s). An accurate knowledge of this form factor is
essential, for example, to calculate the so--called long--distance
contribution to the rare decay $K_L \to \mu^+ \mu^-$. The short--distance
contribution to this process, mediated by loops involving heavy quarks and
massive vector bosons, is sensitive to the CKM matrix element $V_{td}$, but
this contribution cannot be extracted from the accurate experimental
measurement of the partial width unless the long--distance amplitude is
precisely known.

The tree--level rates for several double Dalitz processes have been
published in various forms\cite{kroll,hunt,miyazaki,uy1}. The first 
experimental observation of a double Dalitz decay was published in 
1962\cite{plano,samios2}. A total of 206 examples of the decay $\pzee$ were 
observed by Samios in a sample of some 800,000 bubble chamber photographs. 
Based on the observed angular correlations, Samios was able to exclude the 
possibility that the $\pz$ was a scalar particle (with a CP--conserving 
decay) at the $3.3 \sigma$ confidence level. His measurement of the 
branching ratio for that process remains the only one published to date.

Experimental observations of the much rarer kaon double Dalitz decays began
to appear in the 1990's. Several measurements have been made of the decay
$\klee$\cite{barr1,vagins,akagi1,gu1,barr2,akagi2,ktev_kleeee}, the most
recent of which are based on several hundred observed events. The still 
rarer double Dalitz mode $\klem$ is particularly interesting because it is 
free of complications arising when there are two identical lepton pairs in 
the final state, and because 
it probes only the kinematic region in which one of the virtual photons has 
$q^2 > 4 m_{\mu}^2$. The first example of this decay was reported in 
1996\cite{gu2}. In 2001 the KTeV experiment published a branching ratio
based on a sample of 43 events\cite{ktev_kleemm}; most recently, KTeV has 
reported results from a combined sample of 132 events, including the earlier
43\cite{ktev_klem2}.

Further experimental results on both $\pzee$ and the two kaon decays are 
expected in the near future from the NA48 and KTeV experiments. As the 
statistics available to the experimenters increase, it will be necessary 
to have a more accurate theoretical description of these decays, 
incorporating the significant effects of QED radiative corrections. These 
corrections, discussed in this paper, have a significant impact on the 
extraction of both form factors and angular correlations from 
high--statistics data.

\section{Tree--Level Amplitudes}
The most general CPT invariant interaction governing the transition of a 
spin--zero meson to two photons is 
\begin{equation}
\mathcal{L} = \frac{-i}{4 M} \left[
   \mathcal{F}_P\, \epsilon_{\mu\nu\rho\sigma} 
 + \mathcal{F}_S \left(  g_{\mu\rho} g_{\nu\sigma} 
                       - g_{\mu\sigma} g_{\nu\rho} \right)
 \right] F^{\mu\nu} F^{\rho\sigma} \phi,
\end{equation}
where $\mathcal{F}_P$ and $\mathcal{F}_S$ are dimensionless form factors
for a pseudoscalar and scalar coupling which may be momentum dependent, 
$F^{\mu\nu}$ is the electromagnetic field tensor, and $\phi$ is the field of
a meson of mass $M$. The factor of $-i$ defines a phase convention in which
the form factors will be real if CPT invariance holds and there are no
absorptive decay amplitudes. It will be convenient 
to decompose the couplings into real and imaginary parts, and define them 
(up to an arbitrary overall phase $\Psi$) in terms of a mixing angle $\zeta$, 
and a phase difference $\delta$,
\begin{subequations}
\begin{align}
\mathcal{F}_P & = g_P f_P(k_1^2,k_2^2) e^{i\Psi} 
                = \tilde{g} f_P(k_1^2,k_2^2) \cos{\zeta} e^{i\Psi},\\
\mathcal{F}_S & = g_S f_S(k_1^2,k_2^2) e^{i\Phi} 
                = \tilde{g} f_S(k_1^2,k_2^2) \sin{\zeta} e^{i\delta} e^{i\Psi},
\end{align}
\end{subequations}
where $\tilde{g}^2 = g_P^2 + g_S^2$ and $\delta = \Phi - \Psi$. $f_P$ and 
$f_S$ are functions of the $k^2$ values of the two virtual photons and are 
normalized 
such that $f(0,0) = 1$ for both couplings. In principle it is also possible 
for $\delta$ to depend on $k_1^2$ and $k_2^2$.

\subsection{Two--Photon Decay}%
The two--photon partial width provides information about the coupling
constant $\tilde{g}$ along with some constraints on the mixing angle $\zeta$
and phase difference $\delta$. The matrix element for the decay to two real 
photons with helicities $\lambda_1$ and $\lambda_2$ is
\begin{equation}
\begin{split}
\mel_{\lambda_1 \lambda_2} = & \frac{2}{M} 
  [   \mathcal{F}_P\, \epsilon_{\mu\nu\rho\sigma} \\
  & + \mathcal{F}_S (  g_{\mu\rho} g_{\nu\sigma} 
                     - g_{\mu\sigma} g_{\nu\rho})] 
    k_1^\mu \epsilon_{\lambda_1}^{* \nu} 
    k_2^\rho \epsilon_{\lambda_2}^{* \sigma},
\end{split}
\end{equation}
where $k^\mu$ and $\epsilon^\mu$ are the momentum and polarization of a 
photon. The calculation of the two--photon couplings for photons of
arbitrary mass is carried out in Appendix~\ref{sec:couple}. For real 
photons, the kinematic factors in the couplings reduce to $\lambda = z = 1$ 
and $w = 0$, and the momentum dependent functions $f_P$ and $f_S$ reduce to 
unity. Therefore, one has the following two contributions
\begin{subequations}
\begin{align}
\mel_{++} & = - M \tilde{g} \left(    \sin{\zeta} e^{i \delta}
                                  - i \cos{\zeta} \right),\\
\mel_{--} & = - M \tilde{g} \left(    \sin{\zeta} e^{i \delta}
                                  + i \cos{\zeta} \right).
\end{align}
\end{subequations}
The partial width is obtained by integrating $1/(2M)$ times the squared 
matrix element over the available phase space. The matrix element is a
constant, so the integration results in a factor of $1/(16 \pi)$. The 
partial widths for the two allowed helicity states are then
\begin{subequations}
\begin{align}
\Gamma_{++} & = \frac{M \tilde{g}^2}{32 \pi} 
                \left(1 - 2 \sin{\zeta} \cos{\zeta} \sin{\delta}\right),\\
\Gamma_{--} & = \frac{M \tilde{g}^2}{32 \pi} 
                \left(1 + 2 \sin{\zeta} \cos{\zeta} \sin{\delta}\right).
\end{align}
\end{subequations}
The decay rates to the two final states will be identical if either the 
phase difference between the two couplings is
zero or one of the form factors is zero. In any case, the total rate is equal 
to $\Gamma_{\g\g} = M \tilde{g}^2/(16 \pi)$. An experimental measure of the 
two--photon width then gives a value of $\tilde{g}$,
\begin{equation}
\tilde{g} = \sqrt{16 \pi \Gamma_{\g\g}/M}.
\end{equation}

\subsection{Application to Neutral Pion and Kaon Decays}%
One can compute $\tilde{g}$ for the $\pzgg$, $\klgg$, and $\ksgg$ decays
using the current experimental values\cite{pdg} of the two--photon 
branching ratios along with the meson lifetimes and masses, with the results
\begin{subequations}
\begin{align}
\tilde{g}_0 &= (1.70 \pm 0.06) \cdot 10^{-3},\\
\tilde{g}_L &= (8.75 \pm 0.12) \cdot 10^{-10},\\
\tilde{g}_S &= (1.36 \pm 0.11) \cdot 10^{-9}.
\end{align}
\end{subequations}
It will be useful to define CP--even and CP--odd $\g\g$ states:
\begin{equation}
\vert (\g\g)_\pm \rangle
   = (\vert ++ \rangle \pm \vert -- \rangle)/\sqrt{2}.
\end{equation}
The matrix elements for $\pzgg$ are then given by
\begin{subequations}
\begin{align}
\langle (\g\g)_+ \vert \ T \ \vert \pz \rangle
  &= -i \sqrt{2} M_{\pi} \tilde{g}_0 \sin\zeta_0 e^{i \Phi_0},\\
\langle (\g\g)_- \vert \ T \ \vert \pz \rangle
  &= -\sqrt{2} M_{\pi} \tilde{g}_0 \cos\zeta_0 e^{i \Psi_0}.
\end{align}
\end{subequations}
Similar expressions can be given for the $K_1$ and $K_2$ decay matrix
elements with the 0 subscripts changed to 1 or 2 respectively and $M_{\pi}$
replaced by $M_K$. These expressions may be compared to those in
Ref.~\cite{sehgal} in which CP--violation observables for the $K \to \g\g$
decays were calculated. In that article Sehgal defines
\begin{subequations}
\begin{align}
\langle (\g\g)_+ \vert \ T \ \vert K_1 \rangle 
   &= \phantom{-i} c_e e^{i \rho_e},\\
\langle (\g\g)_- \vert \ T \ \vert K_1 \rangle 
   &= \phantom{-}i c_o e^{i \rho_o},\\
\langle (\g\g)_+ \vert \ T \ \vert K_2 \rangle 
   &= -i d_e e^{i \mu_e},\\
\langle (\g\g)_- \vert \ T \ \vert K_2 \rangle 
   &= \phantom{-i} d_o e^{i \mu_o}.
\end{align}
\end{subequations}
In our notation we have
\begin{subequations}
\begin{align}
c_e &= \phantom{-} \sqrt{2} M_K \tilde{g}_1 \sin\zeta_1,\\
c_o &= -\sqrt{2} M_K \tilde{g}_1 \cos\zeta_1,\\
d_e &= \phantom{-} \sqrt{2} M_K \tilde{g}_2 \sin\zeta_2,\\
d_o &= -\sqrt{2} M_K \tilde{g}_2 \cos\zeta_2.
\end{align}
\end{subequations}
The observable phase differences are given by 
$\delta_1 = \rho_e - \rho_o$ and $\delta_2 = \mu_e - \mu_o$.

The phases $\delta_i$ are observable only if CP is violated in $\g\g$ decays.
As Sehgal noted in Ref.~\cite{sehgal}, the phase $\delta_0$ in $\pz$ decays
should be very close to zero if CPT is conserved because the relevant 
absorptive amplitudes are of order $\alpha^2$. In the case of kaon decays, 
on--shell intermediate states such as $\pi \pi$ couple strongly so that the
phases $\delta_1$ and $\delta_2$ may be large even if CPT is conserved.

\subsection{Four--Lepton Decay}%
\begin{figure}
\includegraphics[bb=120 600 320 740,clip=true]{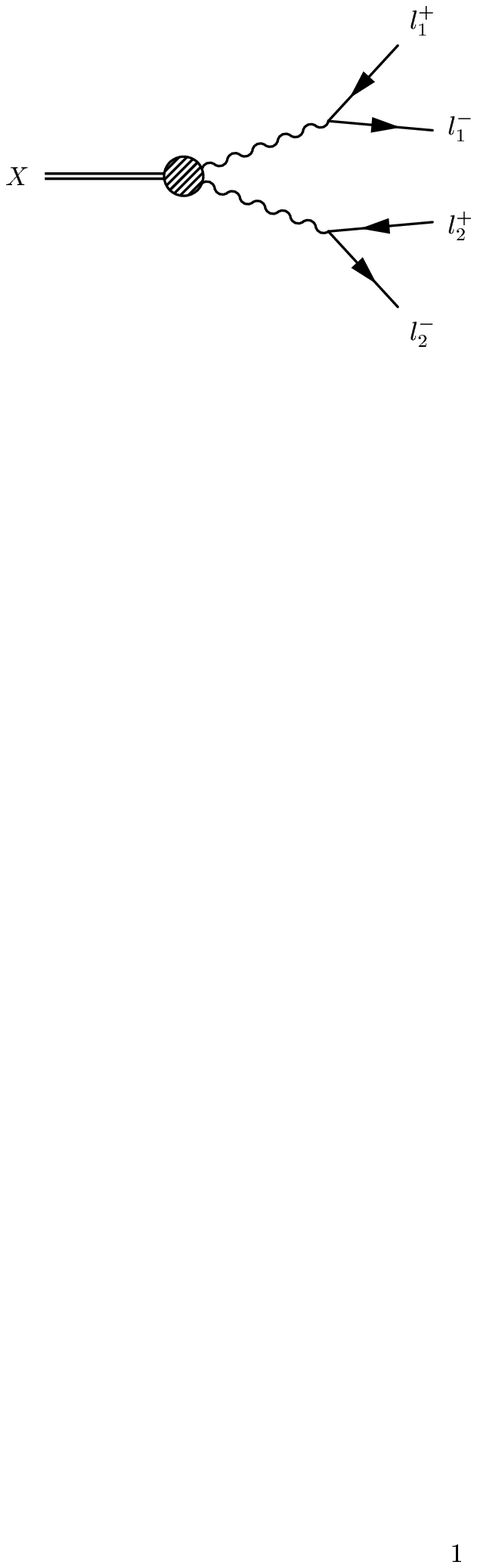}
\caption{\label{fig:feyn01}Double Dalitz Tree--Level Diagram.}
\end{figure}

\begin{figure}
\includegraphics[bb=120 600 320 740,clip=true]{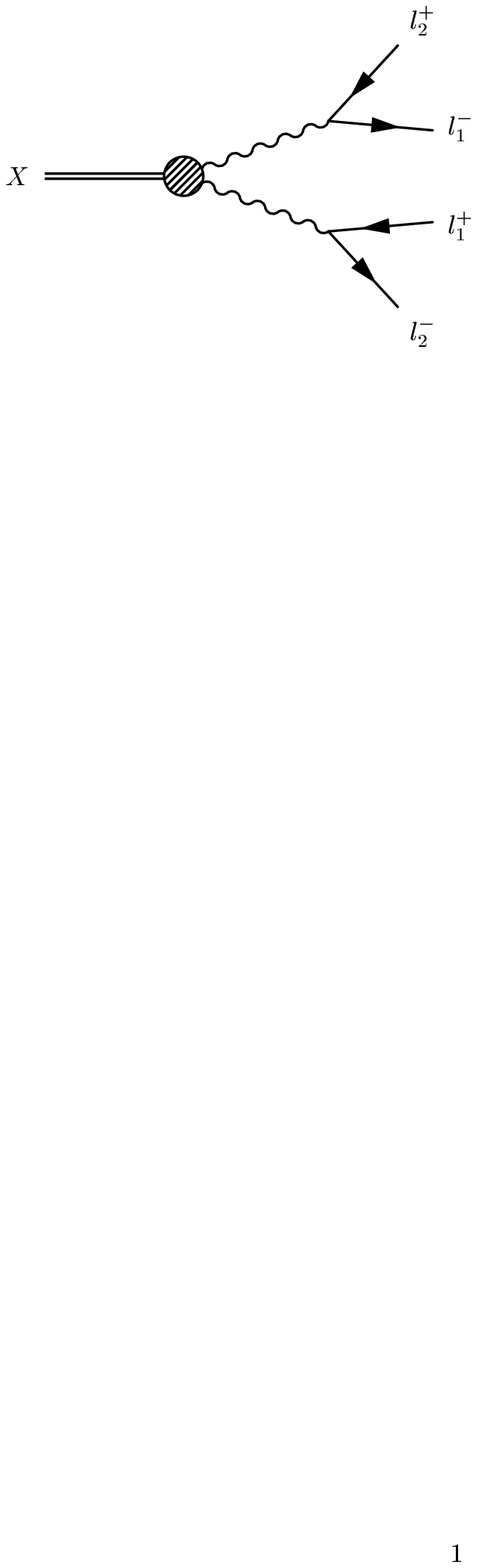}
\caption{\label{fig:feyn02}Double Dalitz Exchange Diagram.}
\end{figure}

While the total two--photon decay rate provides information about the 
constant part of the form factors, the four--lepton rate can be utilized 
to probe both the momentum dependence of the form factors and the mixing 
of the couplings. The decay to four leptons may contain either two pairs 
of identical particles (e.g., $\klee$) or pairs of non--identical particles
(e.g., $\klem$). The tree--level Feynman diagram for the decay is depicted
in Fig.~\ref{fig:feyn01}.
If the final state does contain identical particles, then there is also an
exchange diagram, as shown in Fig.~\ref{fig:feyn02}.
The matrix element is then the sum of the two diagrams 
$\mel = \mel_1 + \mel_2$ and its square is 
$|\mel|^2 = |\mel_1|^2 + |\mel_2|^2 + 2
\real{(\mel_1^* \mel_2)}$. The analysis presented in the rest
of this section applies only to the direct contribution to the double Dalitz
process. Appendix~\ref{sec:interf} gives an explicit expression for the 
interference term.

The matrix element for the direct contribution to the double Dalitz 
process can be written as
\begin{equation}
\mel_1 = H_{\mu\nu\rho\sigma} 
 k_1^\mu  \Pi^{\nu\alpha}   \Gamma_\alpha
 k_2^\rho \Pi^{\sigma\beta} \Gamma_\beta,
\end{equation}
where $H$ is a two--photon coupling given by Eq.~(\ref{eq:coupling}), 
$k$ is a photon momentum, $\Pi$ is the propagator for a photon of momentum $k$
\begin{equation}
\Pi^{\mu\nu}(k) = \frac{i}{k^2} 
 \left(  \sum_\lambda \epsilon_\lambda^{*\mu} \epsilon_\lambda^\nu
       - \hat{k}^\mu \hat{k}^\nu \right),
\end{equation}
and $\Gamma$ is the fermion current for an electron of momentum $q$ and spin
$r$ and positron of momentum $p$ and spin $s$
\begin{equation}
\Gamma^\mu(q,p;r,s) = -ie \bar{u}_r(q) \gamma^\mu v_s(p).
\end{equation}
The sum in the propagator extends over the three helicity states and the
term proportional to $k^\mu k^\nu$ vanishes when contracted with the
current. The matrix element can then be cast as
\begin{equation}
\mel_1 = \sum_{\lambda_1 \lambda_2} 
  \mathcal{H}_{\lambda_1 \lambda_2} L_{\lambda_1} L_{\lambda_2},
\end{equation}
where $\mathcal{H}$ is one of the two--photon couplings given in
Eqs.~(\ref{eq:scouple},\ref{eq:pcouple}). The quantity $L$ contains all
the lepton information and is equal to
$L_\lambda = i \epsilon_\lambda^\mu \Gamma_\mu/k^2$.

The squared matrix element, summed over final state helicities can be
expressed in terms of the five phase space variables $x_{12}$, $x_{34}$, 
$y_{12}$, $y_{34}$, and $\phi$ (defined in Appendix~\ref{sec:kine}). 
Focusing on the $\phi$ dependence, the squared amplitude is
\begin{equation}
\begin{split}
\sum |\mel_1|^2 = \frac{2^8 \pi^2 \alpha^2 \tilde{g}^2}{M^2 w^4}
  &(A \sin^2\phi + B \cos^2\phi\\
  &+ C \sin\phi \cos\phi + D \sin\phi\\
  &+ E \cos\phi + F),
\end{split}
\end{equation}
where 
\begin{subequations}
\begin{align}
A =& w^2 \{ f_P^2 \cos^2\zeta \lambda^2 
             [1 + (1 - \lambda_{12}^2 + y_{12}^2)
                  (1 - \lambda_{34}^2 + y_{34}^2)]\nn\\
   & \quad + f_S^2 \sin^2\zeta z^2
             [2 - \lambda_{12}^2 - \lambda_{34}^2 + y_{12}^2 + y_{34}^2]\},\\
B =& w^2 \{ f_S^2 \sin^2\zeta z^2
             [1 + (1 - \lambda_{12}^2 + y_{12}^2)
                  (1 - \lambda_{34}^2 + y_{34}^2)]\nn\\
   & \quad + f_P^2 \cos^2\zeta \lambda^2 
             [2 - \lambda_{12}^2 - \lambda_{34}^2 + y_{12}^2 + y_{34}^2]\},\\
C =& 2 f_P f_S \sin\zeta \cos\zeta \cos\delta \lambda z w^2\nn\\
   & \quad\quad \times (\lambda_{12}^2 - y_{12}^2)
                       (\lambda_{34}^2 - y_{34}^2),\\
D =& 2 f_P f_S \sin\zeta \cos\zeta \cos\delta \lambda w^3 y_1 y_2\nn\\
   & \quad\quad \times \sqrt{(\lambda_{12}^2 - y_{12}^2)
                             (\lambda_{34}^2 - y_{34}^2)},\\
E =& 2 f_S^2 \sin^2\zeta z w^3 y_{12} y_{34}
     \sqrt{(\lambda_{12}^2 - y_{12}^2)(\lambda_{34}^2 - y_{34}^2)},\\
F =& f_S^2 \sin^2\zeta w^4 (1 - y_{12}^2)(1 - y_{34}^2),
\end{align}
\end{subequations}
where the kinematic variables $w$, $z$, and $\lambda$ are defined in 
Appendix~\ref{sec:kine}.
Terms $A$ and $B$ arise from the diagonal pieces of the transverse part of
the pseudoscalar and scalar couplings while term $F$ is due to the diagonal 
piece of the longitudinal part of the scalar coupling. Term $C$ is the 
interference between the transverse parts of the pseudoscalar and scalar 
couplings while terms $D$ and $E$ are due to interference between the 
longitudinal part of the scalar and the transverse parts of the pseudoscalar 
and scalar couplings respectively.

The partial width is obtained by integrating $1/(2M)$ times the square of 
the matrix element over the eight dimensional phase space (given in 
Appendix~\ref{sec:kine}). The integrals over $y_{ij}$ 
extend from $-\lambda_{ij}$ to $+\lambda_{ij}$ and can be done
analytically. The differential partial width, normalized to the two--photon 
width and integrated over $y_{12}$ and $y_{34}$, is
\begin{equation}
\begin{split}
\frac{1}{\Gamma_{\g\g}}\frac{d\Gamma_1}{d\phi} = 
   \frac{\mathcal{S} \alpha^2}{3 \pi^3} 
   &[  I_1 \cos^2\zeta \sin^2\phi + I_2 \sin^2\zeta \cos^2\phi\\
   & + I_3 \sin\zeta \cos\zeta \cos\delta \sin\phi \cos\phi\\
   & + I_4 \cos^2\zeta + (I_5 + I_6) \sin^2\zeta ],
\end{split}
\end{equation}
where $\mathcal{S}$ is a symmetry factor which is $1/4$ for modes with 
identical particles and 1 otherwise. 
\begin{table}
\caption{Results of numerical integrations over $x_{12}$ and $x_{34}$ assuming
         $f_P = f_S = 1$.\label{tab:ints}}
\begin{ruledtabular}
\begin{tabular}{c|c|c|c|c}
Integral & $\klems$ & $\klmms$              & $\klees$ & $\pzees$ \\ \hline
$I_1$    & 0.10627  & $2.977 \cdot 10^{-6}$ & 14.146   & 7.2287   \\
$I_2$    & 0.11147  & $1.149 \cdot 10^{-5}$ & 14.201   & 7.2838   \\
$I_3$    & 0.21713  & $1.120 \cdot 10^{-5}$ & 28.343   & 14.509   \\
$I_4$    & 0.74203  & $5.499 \cdot 10^{-4}$ & 27.725   & 15.600   \\
$I_5$    & 0.76948  & $1.595 \cdot 10^{-3}$ & 27.809   & 15.684   \\
$I_6$    & 0.01503  & $5.246 \cdot 10^{-4}$ & 0.0556   & 0.0555   
\end{tabular}
\end{ruledtabular}
\end{table}
The $I_i$ factors represent the
integrals over $x_{12}$ and $x_{34}$ given below. The factors $I_1$ and $I_4$
correspond to the pseudoscalar coupling, $I_2$ and $I_5$ are the analogous
terms for the scalar coupling, $I_6$ is the additional longitudinal term in
the scalar coupling, and $I_3$ is the interference term. 
\begin{subequations}
\begin{align}
I_1 =& \frac{2}{3} \iint d x_{12} d x_{34} f_P^2
  \frac{\lambda_{12}^3 \lambda_{34}^3 \lambda^3}{w^2},\\
I_2 =& \frac{2}{3} \iint d x_{12} d x_{34} f_S^2 
  \frac{\lambda_{12}^3 \lambda_{34}^3 \lambda z^2}{w^2},\\
I_3 =& \frac{4}{3} \iint d x_{12} d x_{34} f_P f_S
  \frac{\lambda_{12}^3 \lambda_{34}^3 \lambda^2 z}{w^2},\\
I_4 =& \iint d x_{12} d x_{34} f_P^2
  \frac{\lambda_{12} \lambda_{34} \lambda^3}{w^2} 
  (3 - \lambda_{12}^2 - \lambda_{34}^2),\\
I_5 =& \iint d x_{12} d x_{34} f_S^2 
  \frac{\lambda_{12} \lambda_{34} \lambda z^2}{w^2} 
  (3 - \lambda_{12}^2 - \lambda_{34}^2),\\
I_6 =& \frac{1}{6} \iint d x_{12} d x_{34} f_S^2 
  \lambda_{12} \lambda_{34} \lambda (3 - \lambda_{12}^2)(3 - \lambda_{34}^2).
\end{align}
\end{subequations}
The double integral is performed by first integrating over $x_{34}$ from 
$x_{34}^0$ to $(1-\sqrt{x_{12}})^2$ and then over $x_{12}$ from $x_{12}^0$ to 
$(1-\sqrt{x_{34}^0})^2$, where $x_{ij}^0 = (m_i + m_j)^2/M^2$.
In order to obtain values for these integrals, 
$f_P$ and $f_S$ must first be specified and then the integrals can be done 
numerically. Table~\ref{tab:ints} summarizes the values for the different 
double Dalitz modes assuming that $f_P = f_S = 1$.

The numerical value of $I_5 + I_6$ was found to be several orders of
magnitude larger in Ref.~\cite{uy1}. Extracting $I_6$ from that result
yields a value of 3578.0, compared with our value of 0.01503. This
discrepancy has been traced to the use in Ref.~\cite{uy1} of a scalar
coupling $g_{\mu\rho}g_{\nu\sigma}$, rather than the correctly 
antisymmetrized tensor 
$g_{\mu\rho}g_{\nu\sigma} - g_{\mu\sigma}g_{\nu\rho}$. 
One consequence of the much smaller value we find for $I_6$ is that the
total width for the double Dalitz decay is almost completely insensitive to
the mix of scalar and pseudoscalar couplings (except for the $\klmm$ decay),
in contradistinction to the conclusion of Ref.~\cite{uy1} but in agreement
with the comments near the end of Ref.~\cite{kroll}.

The differential rate can also be expressed in a compact form, suitable for 
experimental fits to the $\phi$ distribution, involving
a constant term, a  CP conserving $\cos{2\phi}$ term, and a CP violating
$\sin{2\phi}$ term
\begin{equation} \label{eq:phi}
\frac{1}{\Gamma_{\g\g}}\frac{d\Gamma_1}{d\phi} = 
  \frac{\alpha^2}{3 \pi^3} 
  R (1 + \kappa_1 \cos{2\phi} + \kappa_2 \sin{2\phi}),
\end{equation}
where
\begin{subequations}
\begin{align}
R        = & \mathcal{S} [(I_1/2 + I_4) \cos^2\zeta \nn\\ 
           & + (I_2/2 + I_5 + I_6) \sin^2\zeta],\\
\kappa_1 = & \mathcal{S} (I_2 \sin^2\zeta - I_1 \cos^2\zeta)/(2 R),\\
\kappa_2 = & \mathcal{S} I_3 \sin\zeta \cos\zeta \cos\delta/(2 R).
\end{align}
\end{subequations}
The values of $R$ and $\kappa_1$ at $\zeta = 0$ and $\zeta = \pi/2$, 
along with the maximum value of $\kappa_2$ and the angle $\zeta_0$ at which 
it takes on that value, are listed in Table~\ref{tab:gamma1}.
\begin{table}
\caption{Coefficients of $\phi$ dependencies for various values of the 
         mixing angle ($\delta=0$).\label{tab:gamma1}}
\begin{ruledtabular}
\begin{tabular}{c|c|c|c|c}
                    & $\klems$ & $\klmms$              & $\klees$ & $\pzees$ \\
 \hline
$R(0)$              & $0.7952$ & $1.379 \cdot 10^{-4}$ & $8.6995$ & $4.8037$ \\
$R(\pi/2)$          & $0.8402$ & $5.313 \cdot 10^{-4}$ & $8.7412$ & $4.8453$ \\
 \hline
$\kappa_1(0)$       & $-0.0668$ & $-0.0027$ & $-0.2033$ & $-0.1881$ \\
$\kappa_1(\pi/2)$   & $+0.0663$ & $+0.0027$ & $+0.2031$ & $+0.1879$ \\ 
 \hline
$\zeta_0$           & $44.32^\circ$ & $26.98^\circ$ 
                    & $44.94^\circ$ & $44.89^\circ$ \\ 
$\kappa_2(\zeta_0)$ & $+0.0664$ & $+0.0026$ & $+0.2031$ & $+0.1880$ 
\end{tabular}
\end{ruledtabular}
\end{table}
As expected, for a pure pseudoscalar decay, the amplitude of the 
$\cos{2\phi}$ term will be negative while the amplitude of the 
$\sin{2\phi}$ vanishes. For a pure scalar decay, the amplitude of the 
$\cos{2\phi}$ term is nearly the same magnitude as in the pseudoscalar 
decay but positive, and the amplitude of $\sin{2\phi}$ again vanishes. 
Depending on the mode, the amplitude for the CP--violating $\sin{2\phi}$ 
term is maximal for values of the mixing angle between $\pi/8$ and $\pi/4$.

Alternatively, we could have integrated over $\phi$ before $x_{12}$ and 
$x_{34}$, in which case we would have
\begin{multline}
\frac{1}{\Gamma_{\g\g}} \frac{d^2 \Gamma_1}{d x_{12} d x_{34}} = 
   \frac{2 \mathcal{S} \alpha^2}{9 \pi^2} 
   \frac{\lambda_{12} \lambda_{34} \lambda}{w^2}
   (3 - \lambda_{12}^2)(3 - \lambda_{34}^2) \\
   \times [  f_P^2 \cos^2\zeta \lambda^2 
           + f_S^2 \sin^2\zeta (\lambda^2 + 3 w^2/2)],
\end{multline}
where we have used $z^2 = \lambda^2 + w^2$. The interference 
term integrates to zero and what remains clearly shows the kinematic 
differences between the contributions of the two couplings.

Going back to Eq.~(\ref{eq:phi}), the final integral over $\phi$ from $0$ to 
$2\pi$ can be performed to get the direct contribution to the four--lepton 
decay rate relative to the two--photon rate
\begin{equation}
\frac{\Gamma_1}{\Gamma_{\g\g}} = 
   \frac{2 \alpha^2}{3 \pi^2} R.
\end{equation}

The total tree--level rate can now be computed for arbitrary form factors for
modes without identical particles in the final state. For the other modes, 
there is the interference between the direct and exchange graphs that must 
be included. The decay rate has the form
\begin{equation} 
\Gamma = \Gamma_1 + \Gamma_2 + \Gamma_{12},
\end{equation}
where, for modes without identical particles $\Gamma_2 = \Gamma_{12} = 0$,
and for modes with identical particles $\Gamma_2 = \Gamma_1$. The expression
given in Appendix~\ref{sec:interf} for the interference term could in
principle be integrated numerically. We choose instead to use a 
Monte Carlo (MC) simulation to integrate the rate and make histograms of the
relevant phase space variables. The decay rates for the various modes, 
broken into diagonal and interference terms, are listed in 
Table~\ref{tab:rate}.
\begin{table}
\caption{The decay rate for pseudoscalar couplings ($\zeta=0$) assuming
         $f_P = f_S = 1$.\label{tab:rate}}
\begin{ruledtabular}
\begin{tabular}{c|c|c|c}
Mode     & $\Gamma_{1+2}/\Gamma_{\g\g}$ 
         & $\Gamma_{12}/\Gamma_{\g\g}$ 
         & $\Gamma/\Gamma_{\g\g}$  \\ \hline
$\klems$ &  $2.859 \cdot 10^{-6}$  
         & $0$                     
         &  $2.859 \cdot 10^{-6}$  \\
$\klmms$ &  $9.914 \cdot 10^{-10}$ 
         & $-0.512 \cdot 10^{-10}$ 
         &  $9.402 \cdot 10^{-10}$ \\
$\klees$ &  $6.256 \cdot 10^{-5}$  
         & $-0.036 \cdot 10^{-5}$  
         &  $6.220 \cdot 10^{-5}$  \\
$\pzees$ &  $3.456 \cdot 10^{-5}$  
         & $-0.036 \cdot 10^{-5}$  
         &  $3.420 \cdot 10^{-5}$  
\end{tabular}
\end{ruledtabular}
\end{table}

Ref.~\cite{miyazaki} included a similar table of values, some of which are
in disagreement with our results. The most significant discrepancy involves
the size of the interference term for the decays $\klee$ and $\pzee$. We find
that the interference in $\klee$ is roughly 9 times smaller than
Ref.~\cite{miyazaki} reports, and that the interference in $\pzee$ is about
4 times smaller. We also differ in the total rate for $\klem$, but the
factor of 2 difference is likely due to a typographical error in the 
previous publication.

The assumption that the form factor is flat contradicts current experimental
findings. The two models that have been used to parameterize the kaon form
factor are the BMS model~\cite{bms} and the DIP model~\cite{dip}. 
The BMS model was originally proposed to describe the coupling in the single 
Dalitz decay $\kleg$ and was therefore written as a function of one $k^2$ only 
\begin{align}
f(x,0) &= \frac{1}{1 - r_\rho x}
      + \frac{C \alpha_{K^*}}{1 - r_{K^*} x}
        \left[\frac{4}{3} - \frac{1}{1 - r_\rho x}\right.\nn\\ 
     &\left. \quad \quad \quad \quad \quad \quad
      - \frac{1}{9}
              \left(\frac{1}{1 - r_\omega x} + \frac{2}{1 - r_\phi x}
        \right) \right].
\end{align}
The quantities $r_i = M^2/M_i^2$ for $M_i$ equal to the $\rho$, $K^*$, 
$\omega$, or $\phi$ meson masses. To apply this model to the double Dalitz
decay, it is assumed that the coupling factors so that 
$f(x_1,x_2) = f(x_1,0) \cdot f(x_2,0)$. In this paper, we will use 
a simplified form of the DIP form factor, which involves only the $\rho$ 
meson and two parameters
\begin{align}
f(x_1,x_2) &= 1 
 + \alpha_{DIP} \left(  \frac{x_1}{x_1 - M_\rho^2/M^2} 
                      + \frac{x_2}{x_2 - M_\rho^2/M^2} \right)\nn\\
 &+ \beta_{DIP} \frac{x_1 x_2}{(x_1 - M_\rho^2/M^2)(x_2 - M_\rho^2/M^2)}.
\end{align}
As will be seen in Appendix~\ref{sec:hf}, the BMS model can be expressed
as a generalized DIP model involving the $\rho$, $\omega$, and $\phi$
vector mesons.

Experimentally, the form factor has
traditionally been linearized in the case of the pion with just a slope
parameter measured, while for the kaon, the BMS model has been
used and values of $\alpha_{K^*}$ quoted. The conversion to the DIP
parameters is easily done, using the world average~\cite{pdg} for the kaon 
we will use $\alpha_{DIP}=-1.5$ and for the pion, $\alpha_{DIP}=-1.0$. There 
is as yet no experimental sensitivity to $\beta_{DIP}$ and so we will use
$\beta_{DIP} = 0$. The effect of using 
the DIP model with these values of $\alpha$ is that the $\pzee$ rate 
increases by less than $0.4\%$, the $\klee$ rate increases by $6.5\%$, the 
$\klem$ rate increases by $56\%$, and the $\klmm$ rate increases by 
$68\%$. It is clear that the assumption of a flat form factor is completely 
invalid for modes containing muons.

\section{Higher Order Processes}
The tree--level double Dalitz process is $\mathcal{O}(\alpha^2)$ since it
contains two electromagnetic vertices. Higher order contributions to the
double Dalitz rate contain one or more internal loops. There are three types
of graphs that contribute at $\mathcal{O}(e^4)$: the vacuum
polarization, the vertex correction, and the 5--point diagram. A
representative diagram from each of these processes is displayed in
Figs.~\ref{fig:feyn03},\ref{fig:feyn04}, and~\ref{fig:feyn05},
respectively. There are two graphs for both the vacuum polarization and the
vertex correction, one for each pair, plus four graphs for the 5--point
function. If there are identical particles in the final state, there are
exchange diagrams and the number of graphs doubles. The interference between 
the tree--level diagram and the one--loop diagrams is
$\mathcal{O}(\alpha^3)$ and therefore contributes to the first order
radiative correction to the double Dalitz rate.

Both the vertex correction and the 5--point graph contain IR
divergences, that is, divergences in the limit that the exchanged photon
energy goes to zero. In order to handle this singular behavior, one must
also consider the radiative double Dalitz decay 
$X \to l^+_1 l^-_1 l^+_2 l^-_2 \g$, in which one of the leptons internally 
radiates a photon. There are two contributions to this process, shown in 
Figs.~\ref{fig:feyn06} and~\ref{fig:feyn07}. The radiative process diverges 
in the opposite manner from the one--loop graphs making the combined decay
rate finite. 

The combined process will be indicated by 
$X \to l^+_1 l^-_1 l^+_2 l^-_2 (\g)$, where the radiated photon may or may
not be detectable. The distinction between the non--radiative and radiative
decays is an experimental issue and ultimately related to the hardware. 
We will use the energy of the radiated photon in the CM frame to 
differentiate between events with a hard photon, $E_\g > \ecut$, and events 
without, $E_\g < \ecut$. The cutoff is chosen such that photons with 
energies below the cutoff can have no significant effect on the 4--lepton
acceptance.
The rate for radiative events with soft photons 
will be added to the rate for non--radiative decays. This contribution is 
also $\mathcal{O}(\alpha^3)$ and therefore must be considered along with 
the one--loop corrections.

The double Dalitz differential rate to second order can therefore be 
expressed as
\begin{equation}
d^5 \Gamma_{\text{rad}} = d^5 \Gamma_{\text{tree}} (1 + \dbrem + \dvirt),
\end{equation}
where $\dbrem$ is the bremsstrahlung contribution due to 
radiative decays with photons below the photon energy cutoff and 
$\dvirt$ is the virtual correction due to the interference
between the tree--level and one--loop diagrams. The virtual correction can
be further decomposed into the contributions from the three one--loop
diagrams 
\begin{equation}
\dvirt = \dvp + \dvc + \dfp,
\end{equation}
where $\dvp$ is the correction from the vacuum polarization diagrams, 
$\dvc$ is the correction from the vertex correction diagrams, and $\dfp$ 
is the correction from the 5--point diagrams. 

\section{Radiative Decays}
The radiative double Dalitz decay will only be considered at tree--level. It 
is straightforward but tedious to write down the expression for the rate.
The two contributions to the rate are shown in Figs.~\ref{fig:feyn06} 
and~\ref{fig:feyn07}. For each process, there exists three additional
diagrams where the photon is radiated off of the other leptons, plus four
exchange graphs if applicable. 

\begin{figure}
\includegraphics[width=3.4in]{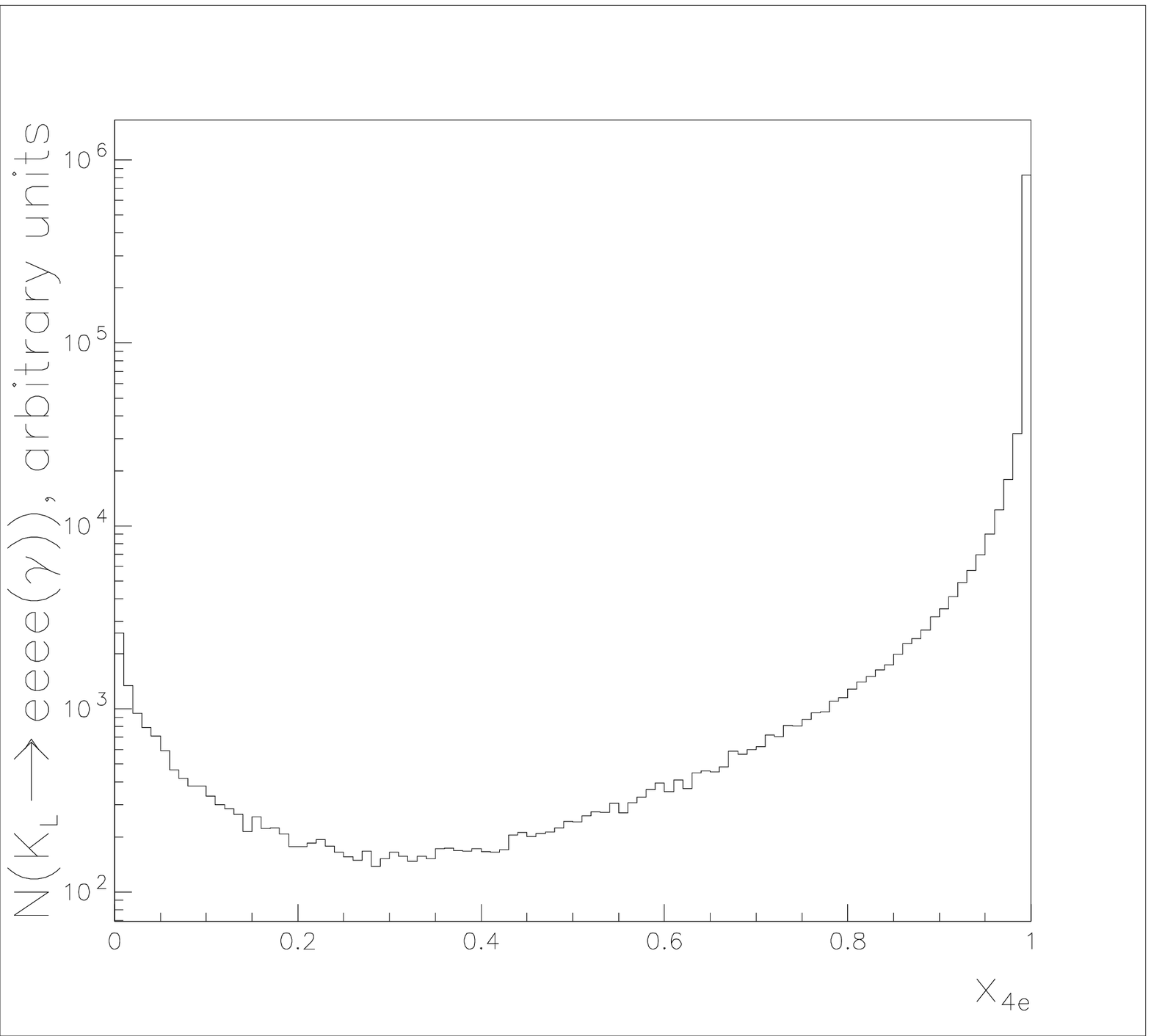}
\caption{\label{fig:x4e}$x_{4e}$ for $\klee$ and $\kleeg$ events.}
\end{figure}

Our results for the radiative decay rates 
use a MC simulation in which we calculate the amplitudes for each helicity 
state using explicit representations of the spinors and polarization vector. 
An photon energy cutoff of $400~\text{keV}$ in the CM frame is used for kaon
decays while for pions, a cutoff of $100~\text{keV}$ is used. It is useful 
to define the quantity $x_{4e} = m_{4e}^2/M^2$, where $m_{4e}$ is the 
reconstructed four--lepton invariant mass, to distinguish between the 
radiative and non--radiative processes. In terms of this variable, the 
cutoff for both kaon and pion decays is at 
$x_{4e}^{\text{cut}} \approx 0.9985$. Fig.~\ref{fig:x4e} shows the 
distribution of $x_{4e}$ for $\klee$ and $\kleeg$ events.
The large peak at $x_{4e} = 1$ is
due to non--radiative events. The part of the distribution which falls away
from the peak at 1 is due to radiated photons from the process of 
Fig.~\ref{fig:feyn06}. The rising part of the distribution near $x_{4e} = 0$
is due to hard Dalitz photons from the process of
Fig.~\ref{fig:feyn07}. Lost due to bin size is the low energy cutoff at 
$E_\g = 400~\text{keV}$ and the high energy cutoff at $x_{4e} = 16 m_e^2/M^2$.

Table~\ref{tab:rad} shows the tree--level radiative decay rates for the four
modes, with no form factor.
\begin{table}
\caption{Tree--level rates for radiative decays including both all radiation
         and only hard radiation such that $x_{4e} < 0.95$, with $f_P=f_S=1$.
         \label{tab:rad}}
\begin{ruledtabular}
\begin{tabular}{l|c|c}
Mode      & $\Gamma_{4l\gamma}(x_{4e}<x_{4e}^{\text{cut}})/\Gamma_{\g\g}$ 
          & $\Gamma_{4l\gamma}(x_{4e}<0.95)/\Gamma_{\g\g}$ \\ \hline
$\pzeegs$ & $6.614(1) \cdot 10^{-6}$ 
          & $2.055(1) \cdot 10^{-6}$  \\
$\kleegs$ & $1.540(1) \cdot 10^{-5}$ 
          & $0.504(1) \cdot 10^{-5}$  \\
$\klemgs$ & $3.279(3) \cdot 10^{-7}$ 
          & $0.873(1) \cdot 10^{-7}$  \\
$\klmmgs$ & $5.634(3) \cdot 10^{-12}$
          & $0.346(1) \cdot 10^{-12}$   
\end{tabular}
\end{ruledtabular}
\end{table}
The rates in the first column include photons of all energies, while
the rates in the second column include only photons with energies large
enough that $x_{4e} < 0.95$. This value of $x_{4e}$ is chosen to closely
match the resolution on the four--lepton mass in current experiments.

\section{Radiative Corrections}
The next four sections will describe the different contributions to the 
radiative corrections to the double Dalitz differential rate. The first 
three contributions are relatively straightforward to determine and we will 
therefore only summarize the relevant formulas. The last contribution, 
coming from the 5--point diagram, is considerably more difficult to calculate. 
In particular, numerical instabilities plague 
the evaluation of the tensor 5--point integrals involving light leptons.
The fourth section, along with much of the 
appendix, will outline our procedure for obtaining this (usually) small but 
non--negligible contribution. We will present the full 5--point diagram
corrections to the differential rate in closed form. In Ref.~\cite{neerven}, 
van~Neerven and Vermaseren reported a numerical integral of the radiative 
corrections to the related two--photon process $\epl \emi \to \epl \emi \pz$ 
but did not present the corrections to the differential cross section.
As we will discuss in Sec.~\ref{sec:results}, the effects of radiative
corrections are much more important compared to form factor effects when
considering double Dalitz decays, because the $q^2$ values in the accessible
phase space are much smaller than those typically probed in two--photon
resonance formation with final state lepton tags.

\subsection{Bremsstrahlung Correction}%
\begin{figure}
\includegraphics[bb=120 600 320 740,clip=true]{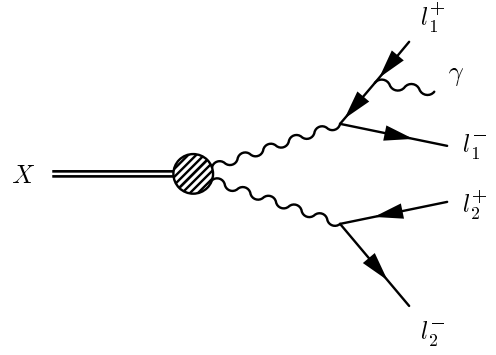}
\caption{\label{fig:feyn06}Radiative Diagram 1.}
\end{figure}

\begin{figure}
\includegraphics[bb=120 600 320 740,clip=true]{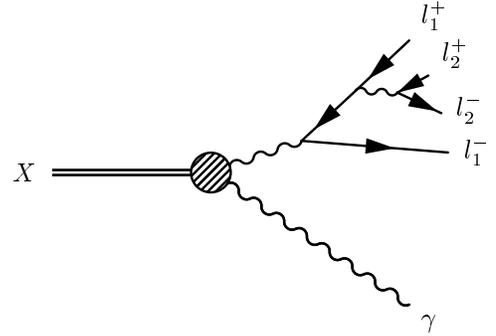}
\caption{\label{fig:feyn07}Radiative Diagram 2.}
\end{figure}

The contribution to the double Dalitz differential decay rate due to the 
soft bremsstrahlung part of the radiative decay is defined as
\begin{equation}
\dbrem(x_{12},x_{34},y_{12},y_{34},\phi) 
  = \frac{d^5 \Gamma_{\text{brem}}/d^5 \Phi}
         {d^5 \Gamma_{\text{tree}}/d^5 \Phi},
\end{equation}
where $d^5 \Gamma_{\text{brem}}/d^5 \Phi$ is the differential decay rate 
for the soft part of the radiative decay integrated over the photon 
momentum with the constraint $E_\g < \ecut$. The full differential rate is
\begin{equation}
d^8 \Gamma_{\text{brem}} = \frac{1}{2M} \sum |\mel_{\text{brem}}|^2
                           d^5 \Phi \frac{d^3 k}{(2 \pi)^3 2 E_k},
\end{equation}
where $d^5 \Phi$ is the four--body phase space differential and 
$\mel_{\text{brem}}$ is the matrix element for the soft bremsstrahlung 
contribution. If the photon energy cutoff is taken small enough, the matrix
element can be approximated as
\begin{equation}
\mel_{\text{brem}} = e \left(
   \frac{p_2 \cdot \epsilon}{p_2 \cdot k}
 + \frac{p_4 \cdot \epsilon}{p_4 \cdot k}
 - \frac{p_1 \cdot \epsilon}{p_1 \cdot k}
 - \frac{p_3 \cdot \epsilon}{p_3 \cdot k} \right) \mel_{\text{tree}},
\end{equation} 
where $\epsilon$ and $k$ are the radiated photon's polarization and 
momentum 4--vectors, respectively. There is one contribution from each of 
the radiative diagrams represented by Fig.~\ref{fig:feyn06}. The other 
type of radiative process (Fig.~\ref{fig:feyn07}) does not contribute in 
this limit. 
If the cutoff is small enough, the lepton momenta can be held fixed while
the photon momentum is integrated out, with the result
\begin{equation}
\dbrem = 4 \pi \alpha 
  \int_0^\ecut \frac{d^3 k}{(2 \pi)^3 2 E_k} \mathcal{B},
\end{equation}
where
\begin{align}
\mathcal{B} &= \sum_\epsilon 
\left| \left(  \frac{p_2^\mu}{p_2 \cdot k} + \frac{p_4^\mu}{p_4 \cdot k}
             - \frac{p_1^\mu}{p_1 \cdot k} - \frac{p_3^\mu}{p_3 \cdot k}
      \right) \epsilon_\mu \right|^2,\\
\begin{split}
   = \frac{2 p_1 \cdot p_2}{(p_1 \cdot k)(p_2 \cdot k)}
   + \frac{2 p_3 \cdot p_4}{(p_3 \cdot k)(p_4 \cdot k)}
   + \frac{2 p_1 \cdot p_4}{(p_1 \cdot k)(p_4 \cdot k)}\\
   + \frac{2 p_2 \cdot p_3}{(p_2 \cdot k)(p_3 \cdot k)}
   - \frac{2 p_2 \cdot p_4}{(p_2 \cdot k)(p_4 \cdot k)}
   - \frac{2 p_1 \cdot p_3}{(p_1 \cdot k)(p_3 \cdot k)}\\
   - \frac{p_1^2}{(p_1 \cdot k)^2}
   - \frac{p_2^2}{(p_2 \cdot k)^2}
   - \frac{p_3^2}{(p_3 \cdot k)^2}
   - \frac{p_4^2}{(p_4 \cdot k)^2}.
\end{split}\nonumber
\end{align}

The correction can be expressed in terms of a sum of ten integrals which 
can be done in closed form
\begin{multline}
\dbrem = 4 \pi \alpha [
    2 I(p_1,p_2) + 2 I(p_3,p_4) + 2 I(p_1,p_4)\\ 
  + 2 I(p_2,p_3) - 2 I(p_2,p_4) - 2 I(p_1,p_3)\\
  - I(p_1,p_1) - I(p_2,p_2) - I(p_3,p_3) - I(p_4,p_4) ],
\end{multline}
where
\begin{equation}
I(k_i,k_j) = \int_0^\ecut 
  \frac{d^3 k}{(2 \pi)^3 2 E_k} \frac{k_i \cdot k_j}
                                     {(k_i \cdot k)(k_j \cdot k)}.
\end{equation}
Each integral yields both a finite part and an IR divergent part which goes as 
$\ln(2 \ecut/\Lambda)$ where $\Lambda$ is the photon mass which will be
taken to zero after the divergent terms are canceled against each other. 
The first two divergent terms will be seen to cancel the divergent parts of 
the vertex correction, the next four cancel divergences in the 5--point 
functions, and the last four cancel the electron self--energy divergences
(which are included in the renormalized vertex function). For $k_i \ne k_j$
\begin{align}
I(k_i,&k_j) = \frac{z_{ij}}{8 \pi^2 \lambda_{ij}} \bigg\{
  \ln{\left(\frac{z_{ij}+\lambda_{ij}}{z_{ij}-\lambda_{ij}}\right)} 
      \ln{\left(\frac{2 \ecut}{\Lambda}\right)}\nn\\
&+ \frac{1}{4} 
  \ln^2{\bigg(\frac{\Omega_i^-}{\Omega_i^+}\bigg)}
 - \frac{1}{4} 
  \ln^2{\bigg(\frac{\Omega_j^-}{\Omega_j^+}\bigg)}\nn\\
   &+ \dilog{\left(1-\frac{\Upsilon_{ij} \Omega_i^+}
                          {x_{ij} \lambda_{ij}}\right)}
    + \dilog{\left(1-\frac{\Upsilon_{ij} \Omega_i^-}
                          {x_{ij} \lambda_{ij}}\right)}\nn\\
   &- \dilog{\left(1-\frac{\Upsilon_{ij} \Omega_j^+}
                          {x_{ij} \lambda_{ij}}\right)}
    - \dilog{\left(1-\frac{\Upsilon_{ij} \Omega_j^-}
                          {x_{ij} \lambda_{ij}}\right)} \bigg\},
\end{align}
where 
\begin{subequations}
\begin{align}
\Omega_i^\pm  &= (1 + \delta_{i,jkl} \pm \lambda_{i,jkl})/2,\\
\Omega_j^\pm  &= (1 + \delta_{j,ikl} \pm \lambda_{j,ikl})/(2 \sigma_{ij}),\\
\Upsilon_{ij} &= \sigma_{ij}(1+\delta_{i,jkl}) - (1+\delta_{j,ikl}),\\
\sigma_{ij}   & = (z_{ij} + \lambda_{ij})/(1 - z_{ij} + \delta_{ij}),
\end{align}
\end{subequations}
and the various $\delta$, $z$, and $\lambda$ symbols are defined in 
Appendix~\ref{sec:kine}. 
For $k_i = k_j$
\begin{equation}
I(k_i,k_i) = \frac{1}{4 \pi^2} \bigg[
  \ln{\left(\frac{2 \ecut}{\Lambda}\right)}
 - \frac{1}{2 \lambda_{ii}} 
  \ln{\left(\frac{1 + \lambda_{ii}}{1 - \lambda_{ii}}\right)} \bigg],
\end{equation}
where $\lambda_{ii}$ is again defined in Appendix~\ref{sec:kine}.

It will be enlightening to extract the IR divergent part of the soft brem
contribution and express it in a way that will make the cancellation
obvious. Collecting terms, one has
\begin{multline}
\dbrem^{IR} = \ln\Lambda \bigg\{
  \frac{2 \alpha}{\pi} \left[1 - \frac{z_{12}}{2 \lambda_{12}}
  \ln{\left(\frac{z_{12}+\lambda_{12}}{z_{12}-\lambda_{12}}\right)}\right]\\
+ \frac{2 \alpha}{\pi} \left[1 - \frac{z_{34}}{2 \lambda_{34}}
  \ln{\left(\frac{z_{34}+\lambda_{34}}{z_{34}-\lambda_{34}}\right)}\right]\\
+ \frac{\alpha}{\pi} \frac{z_{13}}{\lambda_{13}}
  \ln{\left(\frac{z_{13}+\lambda_{13}}{z_{13}-\lambda_{13}}\right)}
- \frac{\alpha}{\pi} \frac{z_{14}}{\lambda_{14}}
  \ln{\left(\frac{z_{14}+\lambda_{14}}{z_{14}-\lambda_{14}}\right)}\\
- \frac{\alpha}{\pi} \frac{z_{23}}{\lambda_{23}}
  \ln{\left(\frac{z_{23}+\lambda_{23}}{z_{23}-\lambda_{23}}\right)}
+ \frac{\alpha}{\pi} \frac{z_{24}}{\lambda_{24}}
  \ln{\left(\frac{z_{24}+\lambda_{24}}{z_{24}-\lambda_{24}}\right)}\bigg\}.
\end{multline}
As will be seen shortly, the first two terms cancel the divergent part of
the vertex correction while the last four cancel the divergent part of the
5--point correction.

\subsection{Virtual Correction}%
As mentioned above, the virtual correction arises from the interference 
between the tree--level and one--loop diagrams. If the full matrix element
is
\begin{equation}
\mel = \mel_{\text{tree}} + \mel_{\text{virt}} 
            + \mathcal{O}(e^6),
\end{equation}
then the squared matrix element to $\mathcal{O}(\alpha^3)$ is
\begin{align}
|\mel_{\text{rad}}|^2 &= |\mel_{\text{tree}}|^2 
  \left[1 
  + \frac{2 \real{(\mel_{\text{tree}}^* \mel_{\text{virt}})}}
         {|\mel_{\text{tree}}|^2}\right],\\
    &= |\mel_{\text{tree}}|^2 
       (1 + \dvirt).
\end{align}
This defines $\dvirt$,
\begin{equation}
\dvirt 
  = \frac{2 \real{(\mel_{\text{tree}}^* \mel_{\text{virt}})}}
         {|\mel_{\text{tree}}|^2}.
\end{equation}
Therefore, we must compute the matrix element for each of the one--loop 
contributions.

\subsubsection{Vacuum Polarization}%
\begin{figure}
\includegraphics[bb=120 600 320 740,clip=true]{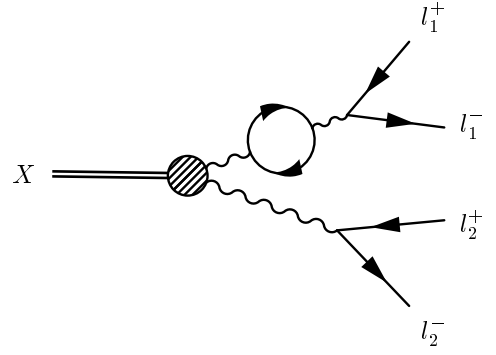}
\caption{\label{fig:feyn03}Vacuum Polarization Diagram.}
\end{figure}

The vacuum polarization process involves higher order corrections to the
photon propagator and is a function of the square of the photon momentum, or
the $x$ of that pair. There is one contribution for each photon propagator.
One contribution is shown in Fig.~\ref{fig:feyn03}.
The vacuum polarization diagram is IR finite but UV
divergent. The divergence can be handled by renormalization of the photon
wavefunction. The vacuum polarization matrix element can be written as the
tree--level matrix element times the renormalized polarization insertion 
\begin{equation}
\mel_{\text{vp}} = \mel_{\text{tree}} \sum_l \Pi_l(x_{ij}),
\end{equation}
where the sum is over lepton species in the loop 
and the renormalized polarization insertion is
\begin{multline}
\Pi_l(x_{ij}) = \frac{2 \alpha}{\pi} \int_0^1 d z\ z (1-z)\\
   \times \ln\left[1 - z (1-z) x_{ij} M^2/m_l^2 - i \epsilon\right],
\end{multline}
where $m_l$ is the mass of the lepton in the loop. The integration depends 
on the size of $x_{ij}$ compared to $m_l^2/M^2$, such that
\begin{multline}
\Pi_l(x_{ij}) = -\frac{\alpha}{3 \pi} \bigg\{\frac{8}{3} - \beta_{ij}^2\\ 
   + \frac{\beta_{ij}}{2} (3 - \beta_{ij}^2)
  \left[\ln\left(\frac{1-\beta_{ij}}{1+\beta_{ij}}\right) + i\pi\right]\bigg\},
\end{multline}
for $x_{ij} > 4 m_l^2/M^2$, while
\begin{multline}
\Pi_l(x_{ij}) = -\frac{\alpha}{3 \pi} \bigg\{\frac{8}{3} + \rho_{ij}^2\\ 
   - \frac{\rho_{ij}}{2} (3 + \rho_{ij}^2)
  \left[\pi - 2 \tan^{-1} \rho_{ij}\right]\bigg\},
\end{multline}
for $x_{ij} < 4 m_l^2/M^2$. The functions $\rho$ and $\beta$ are related to
$\lambda$ as defined in Appendix~\ref{sec:kine} but are functions of the
loop mass
\begin{subequations}
\begin{align}
\beta_{ij} & = \sqrt{1 - 4 m_l^2/(x_{ij} M^2)},\\
\rho_{ij} & = \sqrt{4 m_l^2/(x_{ij} M^2) - 1}.
\end{align}
\end{subequations}

The correction then is
\begin{equation}
\dvp = 2 \sum_g \left[\sum_{l_g} \real{\Pi_{l_g}(x_{ij})}\right],
\end{equation}
where the first sum is over the number of vacuum polarization graphs and the 
second sum is over the possible lepton species in the loop.

\subsubsection{Vertex Function}%
\begin{figure}
\includegraphics[bb=120 600 320 740,clip=true]{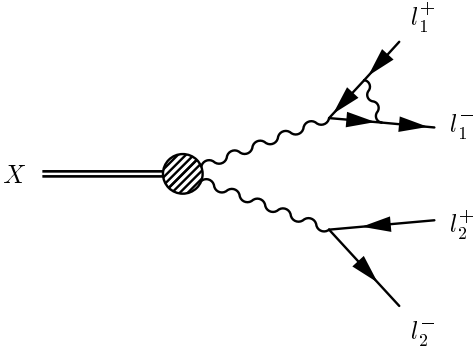}
\caption{\label{fig:feyn04}Vertex Correction Diagram.}
\end{figure}

The vertex function involves higher order corrections to the QED vertex and
is a function of the momenta of the pair. One contribution is shown
in Fig.~\ref{fig:feyn04}. 
\begin{figure}
\includegraphics[bb=120 600 320 740,clip=true]{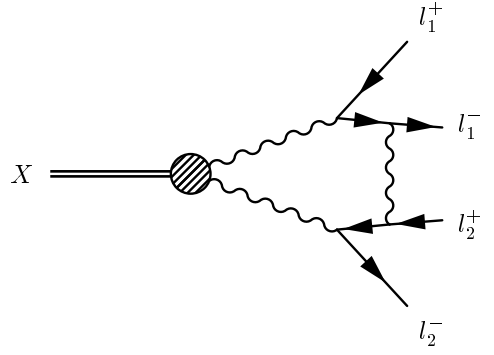}
\caption{\label{fig:feyn05}5--Point Diagram.}
\end{figure}
The vertex correction contains both UV and IR divergences. 
We will also include the self--energy correction to the lepton lines which 
also are UV and IR divergent. Both UV divergences will be handled 
simultaneously by renormalization of the electromagnetic coupling and the 
lepton wavefunction while the IR divergent part will cancel the IR 
divergence in the soft bremsstrahlung correction. The matrix element 
corresponding to one of the diagrams is
\begin{equation}
\mel_{\text{vc}} = \mel_{\text{tree}} \mathcal{V}(x_{ij},y_{ij}),
\end{equation}
where 
\begin{equation}
\mathcal{V}(x_{ij},y_{ij}) = F_1(x_{ij}) + F_2(x_{ij}) 
  \left[\frac{2}{2 - \lambda_{ij}^2 + y_{ij}^2}\right].
\end{equation}
$F_1$ and $F_2$ are vertex form factors defined by
\begin{align}
F_1(&x_{ij}) =  \frac{\alpha}{\pi} \Bigg(
   \left\{1 - \frac{z_{ij}}{2 \lambda_{ij}} 
     \left[\ln{\left(\frac{z_{ij}+\lambda_{ij}}{z_{ij}-\lambda_{ij}}\right)} 
          - 2i\pi\right]\right\} 
   \ln\frac{m_i}{\Lambda}\nn\\
 &-1 + \frac{1 + 2 \lambda_{ij}^2}{8 \lambda_{ij}} 
     \left[\ln{\left(\frac{z_{ij}+\lambda_{ij}}{z_{ij}-\lambda_{ij}}\right)}
          - 2i\pi\right]\nn\\
 &- \frac{z_{ij}}{\lambda_{ij}} 
   \left[\dilog\left(\frac{2\lambda_{ij}}{1+\lambda_{ij}}\right) 
    + \frac{1}{16}
      \ln^2{\left(\frac{z_{ij}+\lambda_{ij}}{z_{ij}-\lambda_{ij}}\right)} 
    - \frac{\pi^2}{2}\right] \nn\\
 &+ \frac{i \pi}{2} \frac{z_{ij}}{\lambda_{ij}} 
   \ln\left(\frac{2 \lambda_{ij}^2}{1 - z_{ij}}\right)\Bigg),\\
F_2(&x_{ij}) = -\frac{\alpha}{\pi} \frac{1 - z_{ij}}{4 \lambda_{ij}} 
   \left[\ln{\left(\frac{z_{ij}+\lambda_{ij}}{z_{ij}-\lambda_{ij}}\right)} 
         - 2i\pi\right],
\end{align}
where $z_{ij}$ and $\lambda_{ij}$ are defined in Appendix~\ref{sec:kine} and
$\Lambda$ is the photon mass. 

The correction is then just
\begin{equation}
\dvc = 2 \sum_g \real{\mathcal{V}(x_{ij},y_{ij})},
\end{equation}
where the sum over $g$ is over the number of vertex function graphs.

The IR divergent part is contained within $F_1$. Twice the real part of the
$\ln\Lambda$ term is exactly what is necessary to cancel the first two
divergent terms (one for each vertex correction graph) in the soft brem 
contribution. The final four divergent soft brem terms will have to wait 
for the last piece of the virtual correction, the 5--point diagram.

\subsubsection{5--Point Diagram}%
There are four distinct 5--point diagrams that contribute to the direct
process, plus four more if there is an exchange process. The diagram shown
in Fig.~\ref{fig:feyn05} contains a photon exchanged between $l^-_1$ and 
$l^+_2$. The matrix element for that graph is
\begin{widetext}
\begin{multline}
\mel_{\text{5p}}^1 = \int \frac{d^4 t}{(2 \pi)^4} \frac{2}{M}
 \left[  \mathcal{F}_P \epsilon_{\mu\nu\rho\sigma} 
       + \mathcal{F}_S (  g_{\mu\rho}g_{\nu\sigma} 
                        - g_{\mu\sigma}g_{\nu\rho})\right]
        (p_{12} + t)^\mu (p_{34} - t)^\rho \\
 \times \left[\frac{-i g^{\delta\eta}}
                   {t^2 - \lambda^2 + i\epsilon}\right]
        \left[\frac{-i g^{\alpha\nu}}
                   {(p_{12}+t)^2 - \lambda^2 + i\epsilon}\right]
        \left[\frac{-i g^{\beta\sigma}}
                   {(p_{34}-t)^2 - \lambda^2 + i\epsilon}\right] \\
 \times \bar{u}(p_2) (-i e\gamma_\delta) 
        \left(\frac{i}{\rlap/{t} + \rlap/{p_2} - m_1 + i\epsilon}\right)
        (-i e\gamma_\alpha) v(p_1)
        \bar{u}(p_4) (-i e\gamma_\beta) 
        \left(\frac{i}{\rlap/{t} - \rlap/{p_3} - m_2 + i\epsilon}\right)
        (-i e\gamma_\eta) v(p_3),
\end{multline}
where $t$ is the loop momentum and $p_1$, $p_2$, $p_3$, and $p_4$ are the
momenta of $l^+_1$, $l^-_1$, $l^+_2$, and $l^-_2$, respectively. 
This can be re--expressed as
\begin{align}
\mel_{\text{5p}}^1& = -\frac{2 i e^4}{M} \bigg(
    \mathcal{F}_P\ \epsilon_{\mu\rho\nu\sigma} \big\{
   - p_{12}^\mu p_{34}^\rho \big[  
         I_{50}               A^{\nu\sigma} 
       + I_{51}^{\alpha}      B^{\nu\sigma}_{\ \ \alpha}
       + I_{52}^{\alpha\beta} C^{\nu\sigma}_{\ \ \alpha\beta}
     \big]
   + p_5^{\mu} \big[
         I_{51}^{\rho}            A^{\nu\sigma} 
       + I_{52}^{\rho\alpha}      B^{\nu\sigma}_{\ \ \alpha}
       + I_{53}^{\rho\alpha\beta} C^{\nu\sigma}_{\ \ \alpha\beta}
     \big]\big\}\nn\\
  & + \mathcal{F}_S \big\{
    [(p_{12} \cdot p_{34}) g^{\mu \nu} - p_5^\mu p_5^\nu] \big[
         I_{50}               A_{\mu\nu} 
       + I_{51}^{\alpha}      B_{\mu\nu\alpha}
       + I_{52}^{\alpha\beta} C_{\mu\nu\alpha\beta}
     \big]
   + (p_{34} - p_{12})_\rho g_{\mu \nu} \big[
         I_{51}^{\rho}            A^{\mu\nu} 
       + I_{52}^{\rho\alpha}      B^{\mu\nu}_{\ \ \alpha}
       + I_{53}^{\rho\alpha\beta} C^{\mu\nu}_{\ \ \alpha\beta} 
     \big]\nn\\
 & \quad \quad \quad 
   - g_{\mu\nu} \big[  
       (I_{52})^{\rho}_{\rho}            A^{\mu\nu} 
     + (I_{53})^{\rho}_{\alpha\rho}      B^{\alpha\mu\nu}
     + (I_{54})^{\rho}_{\alpha\beta\rho} C^{\alpha\beta\mu\nu}
    \big] \big\} \bigg),\label{eq:5pmat}
\end{align}
\end{widetext}
where $A$, $B$, and $C$ are combinations of spinors and gamma matrices
and $p_5 = p_1+p_2+p_3+p_4$.
The factors of $I_i$ are integrals over the loop momentum. There are three
basic integral forms from which all the other may be obtained. The notation 
for the integrals has the following meaning: the first digit in the subscript 
refers to the number of denominators and the second refers the number of 
powers of the loop momentum appearing in the integral. The 5--point
integrals are defined in Appendix~\ref{sec:hf} as a function of four
4--vector arguments $k_1$, $k_2$, $k_3$, $k_4$.  

For the diagram shown in Fig.~\ref{fig:feyn05}, the arguments of the 
5--point integral functions should take on the values
\begin{subequations}
\begin{align}
k_1 &= -p_2, & k_2 &= -p_1,\\
k_3 &= p_5,  & k_4 &= -p_4,
\end{align}
\end{subequations}
so we can write
\begin{equation}
I_{50} = I_{50}(-p_2, -p_1, p_5, -p_4)
\end{equation}
for the scalar integral, and analogous expressions for the higher-rank
tensor integrals.
The spinor terms for this diagram are
\begin{subequations}
\begin{align}
A^{\mu\nu} &= -4 (p_2 \cdot p_3) \bar{u}(p_2) \gamma^\mu v(p_1)
                                 \bar{u}(p_4) \gamma^\nu v(p_3),\\
B^{\mu\nu\rho} &= 2 [\bar{u}(p_2) \gamma^\nu v(p_1)
                     \bar{u}(p_4) \gamma^\rho \gamma^\mu \rlap/p_2 v(p_3)
 \nonumber\\   & \quad - 
                     \bar{u}(p_2) \rlap/p_3 \gamma^\mu \gamma^\nu v(p_1)
                     \bar{u}(p_4) \gamma^\rho v(p_3)],\\
C^{\mu\nu\rho\sigma} &= 
            \bar{u}(p_2) \gamma^\eta \gamma^\mu \gamma^\rho v(p_1)
            \bar{u}(p_4) \gamma^\sigma \gamma^\nu \gamma_\eta v(p_3).
\end{align}
\end{subequations}

The spinor terms for the diagram containing a photon exchanged between $l^-_1$
and $l^-_2$ are
\begin{subequations}
\begin{align}
A^{\mu\nu} &=  4 (p_2 \cdot p_4) \bar{u}(p_2) \gamma^\mu v(p_1)
                                 \bar{u}(p_4) \gamma^\nu v(p_3),\\
B^{\mu\nu\rho} &= 2 [\bar{u}(p_2) \rlap/p_4 \gamma^\mu \gamma^\nu v(p_1)
                     \bar{u}(p_4) \gamma^\rho v(p_3)
 \nonumber\\   & \quad - 
                     \bar{u}(p_2) \gamma^\nu  v(p_1)
                     \bar{u}(p_4) \rlap/p_2 \gamma^\mu \gamma^\rho v(p_3)],\\
C^{\mu\nu\rho\sigma} &= 
          - \bar{u}(p_2) \gamma^\eta \gamma^\mu \gamma^\rho v(p_1)
            \bar{u}(p_4) \gamma_\eta \gamma^\nu \gamma^\sigma v(p_3),
\end{align}
\end{subequations}
and the scalar integral for this diagram is 
\begin{equation}
I_{50} = I_{50}(-p_2, -p_1, p_5, -p_3).
\end{equation}

The spinor terms for the diagram containing a photon exchanged between $l^+_1$
and $l^+_2$ are
\begin{subequations}
\begin{align}
A^{\mu\nu} &=  4 (p_1 \cdot p_3) \bar{u}(p_2) \gamma^\mu v(p_1)
                                 \bar{u}(p_4) \gamma^\nu v(p_3),\\
B^{\mu\nu\rho} &= 2 [\bar{u}(p_2) \gamma^\nu \gamma^\mu \rlap/p_3 v(p_1)
                     \bar{u}(p_4) \gamma^\rho v(p_3)
 \nonumber\\   & \quad - 
                     \bar{u}(p_2) \gamma^\nu v(p_1)
                     \bar{u}(p_4) \gamma^\rho \gamma^\mu \rlap/p_1 v(p_3)],\\
C^{\mu\nu\rho\sigma} &= 
          - \bar{u}(p_2) \gamma^\rho \gamma^\mu \gamma^\eta v(p_1)
            \bar{u}(p_4) \gamma^\sigma \gamma^\nu \gamma_\eta v(p_3),
\end{align}
\end{subequations}
and the scalar integral for this diagram is 
\begin{equation}
I_{50} = I_{50}(-p_1, -p_2, p_5, -p_4).
\end{equation}

The spinor terms for the diagram containing a photon exchanged between $l^+_1$
and $l^-_2$ are
\begin{subequations}
\begin{align}
A^{\mu\nu} &= -4 (p_1 \cdot p_4) \bar{u}(p_2) \gamma^\mu v(p_1)
                                 \bar{u}(p_4) \gamma^\nu v(p_3),\\
B^{\mu\nu\rho} &= 2 [\bar{u}(p_2) \gamma^\nu v(p_1)
                     \bar{u}(p_4) \rlap/p_1 \gamma^\mu \gamma^\rho v(p_3)
 \nonumber\\   & \quad - 
                     \bar{u}(p_2) \gamma^\nu \gamma^\mu \rlap/p_4 v(p_1)
                     \bar{u}(p_4) \gamma^\rho v(p_3)],\\
C^{\mu\nu\rho\sigma} &= 
            \bar{u}(p_2) \gamma^\rho \gamma^\mu \gamma^\eta v(p_1)
            \bar{u}(p_4) \gamma_\eta \gamma^\nu \gamma^\sigma v(p_3),
\end{align}
\end{subequations}
and the scalar integral for this diagram is 
\begin{equation}
I_{50} = I_{50}(-p_1, -p_2, p_5, -p_3).
\end{equation}

The tensors $A$, $B$, and $C$ are computed for a given helicity combination 
and combined with the integrals to yield the matrix element for that
helicity state. The correction then involves a sum over the sixteen
possible final states
\begin{equation}
\dfp 
= \frac{\sum_{\lambda=1}^{16} 2 \real{\left[\mel_{\text{tree}}^*(\lambda) 
                                            \mel_{\text{5p}}(\lambda)\right]}}
       {\sum_{\lambda=1}^{16} \left|\mel_{\text{tree}}(\lambda)\right|^2},
\end{equation}
where $\lambda$ here refers to the helicity state and
\begin{align}
\mel_{\text{tree}}(\lambda) & = \sum_{g} \mel_{\text{tree}}^g(\lambda),\\
\mel_{\text{5p}}(\lambda)   & = \sum_{g} \mel_{\text{5p}}^g(\lambda).
\end{align}
The sums here are over the number of graphs for each process.

The IR divergent part of the 5--point correction is most easily isolated 
by looking at the 5--point matrix element in the IR limit. All terms 
involving tensor integrals vanish leaving only the $I_{50}$ term. The 
divergent part of $I_{50}$ is due to the two divergent box integrals,
$I_{40}^{(3)}$ and $I_{40}^{(4)}$. The relevant terms in the scalar 5--point
function for Fig.~\ref{fig:feyn05}, in terms of the divergent 3--point 
function, are
\begin{align}
I_{50}^{IR} & = -\frac{1}{2} 
  \left(  \frac{\sum_j S_{3j}^{-1}}{p_{34}^2 - M_2^2}
        + \frac{\sum_j S_{4j}^{-1}}{p_{12}^2 - M_1^2} \right) I_{IR},\nn\\
            & = I_{IR}/\left[(p_{12}^2 - M_1^2)(p_{34}^2 - M_2^2)\right].
\end{align}
Extracting the $\ln\Lambda$ piece of $I_{IR}$, $I_{50}^{IR}$ can be written
as
\begin{multline}
I_{50}^{IR} = \frac{-i}{16 \pi^2 \lambda_{23} p_{23}^2 
                       (p_{12}^2 - M_1^2) (p_{34}^2 - M_2^2)}\\ \times
 \left[\ln{\left(\frac{z_{23} + \lambda_{23}}{z_{23} - \lambda_{23}}\right)} 
       - 2 i \pi \right] \ln\frac{\Lambda}{\mu},
\end{multline}
where $\mu$ is a kinematic function with dimensions of mass which is 
independent of $\Lambda$.
The divergent part of the 5--point matrix element, dropping the finite
term involving $\mu$, is proportional to the tree--level matrix element
\begin{equation}
\mel_{\text{5p}}^{IR} = \mel_{\text{tree}} \frac{\alpha}{2 \pi}
  \frac{z_{23}}{\lambda_{23}} 
 \left[\ln{\left(\frac{z_{23} + \lambda_{23}}{z_{23} - \lambda_{23}}\right)} 
       - 2 i \pi \right] \ln\Lambda.
\end{equation}
The IR divergent part of the 5--point correction coming from all four
diagrams is
\begin{equation}
\dfp^{IR} = \sum_g -s_{ij} \frac{\alpha}{\pi} \frac{z_{ij}}{\lambda_{ij}} 
   \ln{\left(\frac{z_{ij}+\lambda_{ij}}{z_{ij}-\lambda_{ij}}\right)}
   \ln\Lambda,
\end{equation}
where $s_{ij}$ is the product of the sign of the charges of $p_i$ and
$p_j$ and the sum is over the four diagrams. Again, it can be seen that this 
is the necessary form to cancel the remaining four divergent brem terms.

\section{MC Simulation Results\label{sec:results}}
\begin{figure}
\includegraphics[width=3.4in]{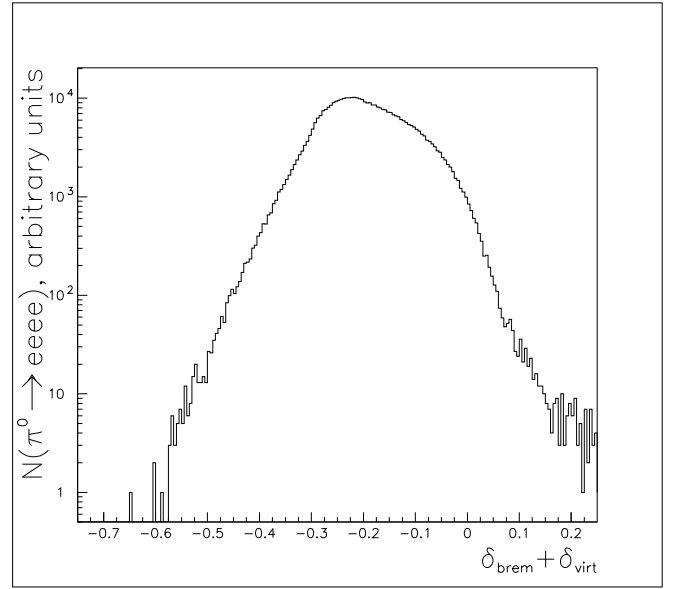}
\caption{Distribution of the total radiative correction 
         for $\pzees$ events with an IR cutoff of 
         $x_{4e}^{\text{cut}}=0.9985$, with $f_P=f_S=1$.\label{fig:delta}}
\end{figure}

The inclusion of the radiative corrections impacts both the differential
rate and the total rate. The total correction to the differential $\pzee$ 
rate is shown in Fig.~\ref{fig:delta}.
The average size of the correction factor for the four different modes 
is shown in Table~\ref{tab:avgd}.
\begin{table}
\caption{Average size of radiative correction to the differential rate with 
         an IR cutoff of $x_{4e}^{\text{cut}}=0.9985$, with $f_P=f_S=1$.
         \label{tab:avgd}}
\begin{ruledtabular}
\begin{tabular}{l|c|c|c|c}
               & $\pzees$  & $\klees$  & $\klems$  & $\klmms$ \\ \hline
$\bar{\delta}$ & $-0.1948$ & $-0.2618$ & $-0.0788$ & $+0.0805$
\end{tabular}
\end{ruledtabular}
\end{table}
The total rate for the combined 4--lepton plus photon process is 
independent of IR cutoff. Table~\ref{tab:rates} summarizes the tree--level 
rate and the rate for the combined, cutoff independent process, divided into
the rate including all radiation and the rate including only soft radiation
($x_{4e} > 0.95$), all with $f_P=f_S=1$. 
\begin{table}
\caption{Summary of tree--level 4--lepton rate and combined radiatively 
         corrected 4--lepton plus photon rate, including the rate for all 
         $x_{4e}$ and the rate for $x_{4e}>0.95$, using $f_P=f_S=1$.
         \label{tab:rates}}
\begin{ruledtabular}
\begin{tabular}{l|c|c|c}
         &     
         & 
\multicolumn{2}{c}{$\Gamma_{4l(\g)}/\Gamma_{\g\g}$} \\ \cline{3-4}
\raisebox{1.5ex}[0pt]{Mode}     &
\raisebox{1.5ex}[0pt]{$\Gamma_{4l}^{tree}/\Gamma_{\g\g}$} &
         all $x_{4e}$ &
         $x_{4e} > 0.95$ \\ \hline
$\pzees$ & $3.421(4) \cdot 10^{-5}$ 
         & $3.536(4) \cdot 10^{-5}$
         & $3.331(4) \cdot 10^{-5}$ \\
$\klees$ & $6.222(5) \cdot 10^{-5}$ 
         & $6.406(4) \cdot 10^{-5}$ 
         & $5.903(4) \cdot 10^{-5}$ \\
$\klems$ & $2.858(1) \cdot 10^{-6}$ 
         & $2.996(3) \cdot 10^{-6}$
         & $2.909(3) \cdot 10^{-6}$ \\
$\klmms$ & $0.941(1) \cdot 10^{-9}$ 
         & $1.026(1) \cdot 10^{-9}$ 
         & $1.025(1) \cdot 10^{-9}$ 
\end{tabular}
\end{ruledtabular}
\end{table}
It is the last column which should most accurately predict the observed 
non--radiative 4--lepton rate. It is seen that the non--radiative rate is 
smaller than the tree--level rate for both 4--electron modes while it is 
larger for the modes with muons.

The probability of radiation can now be computed as the ratio of the
radiative rate to the combined rate. Table~\ref{tab:radprob} lists the
probability of radiating a photon ($x_{4e} < 0.9985$) along with the
probability of radiating a hard photon ($x_{4e} < 0.95$) for each of the
four modes.
\begin{table}
\caption{Probability of radiation ($x_{4e} < 0.9985$) and probability of
         hard radiation ($x_{4e} < 0.95$), defined as 
         $P = \Gamma_{4l\g}/\Gamma_{4l(\g)}$.\label{tab:radprob}}
\begin{ruledtabular}
\begin{tabular}{l|c|c}
Mode     & $P(x_{4e} < 0.9985)$ & $P(x_{4e} < 0.95)$ \\ \hline
$\pzeegcs$ & 0.187 & 0.058 \\ 
$\kleegcs$ & 0.240 & 0.079 \\
$\klemgcs$ & 0.109 & 0.029 \\ 
$\klmmgcs$ & 0.006 & 0.0003 
\end{tabular}
\end{ruledtabular}
\end{table}
The probability is highest for $\kleegc$ where the $x$ values can be the
smallest. The probabilities for $\klemgc$ are slightly less than half of what
they are for the four electron mode, and in $\klmmgc$, there is very little
radiation. 

The effect of the radiative corrections on the differential rate can be
observed in the distributions of the five phase space variables. The
statistics in the following plots reflect the amount of CPU time dedicated
to each mode. While the calculation of the radiative corrections is 
CPU--intensive, it is actually the generation of the radiative decays that 
takes the most time.

For the
modes with identical leptons, it is useful to adopt a method of pairing the
electrons with the positrons in order to study the dilepton mass 
distributions. We choose to use the pairing for which the
product of $x$'s is minimized. It is this pairing that will contribute the
most to the matrix element in general. Therefore, $x_a$ and $x_b$ are 
the $x$'s belonging to this pairing, with the additional requirement 
that $x_a < x_b$. In addition, $y_a$ is the $y$ variable defined in the
$a$--pair CM, and $y_b$ is the same quantity in the $b$--pair CM. And
lastly, $\phi_{ab}$ is the angle between the planes of the $a$--pair and
$b$--pair in the overall CM. 

\begin{figure}
\includegraphics[width=3.4in]{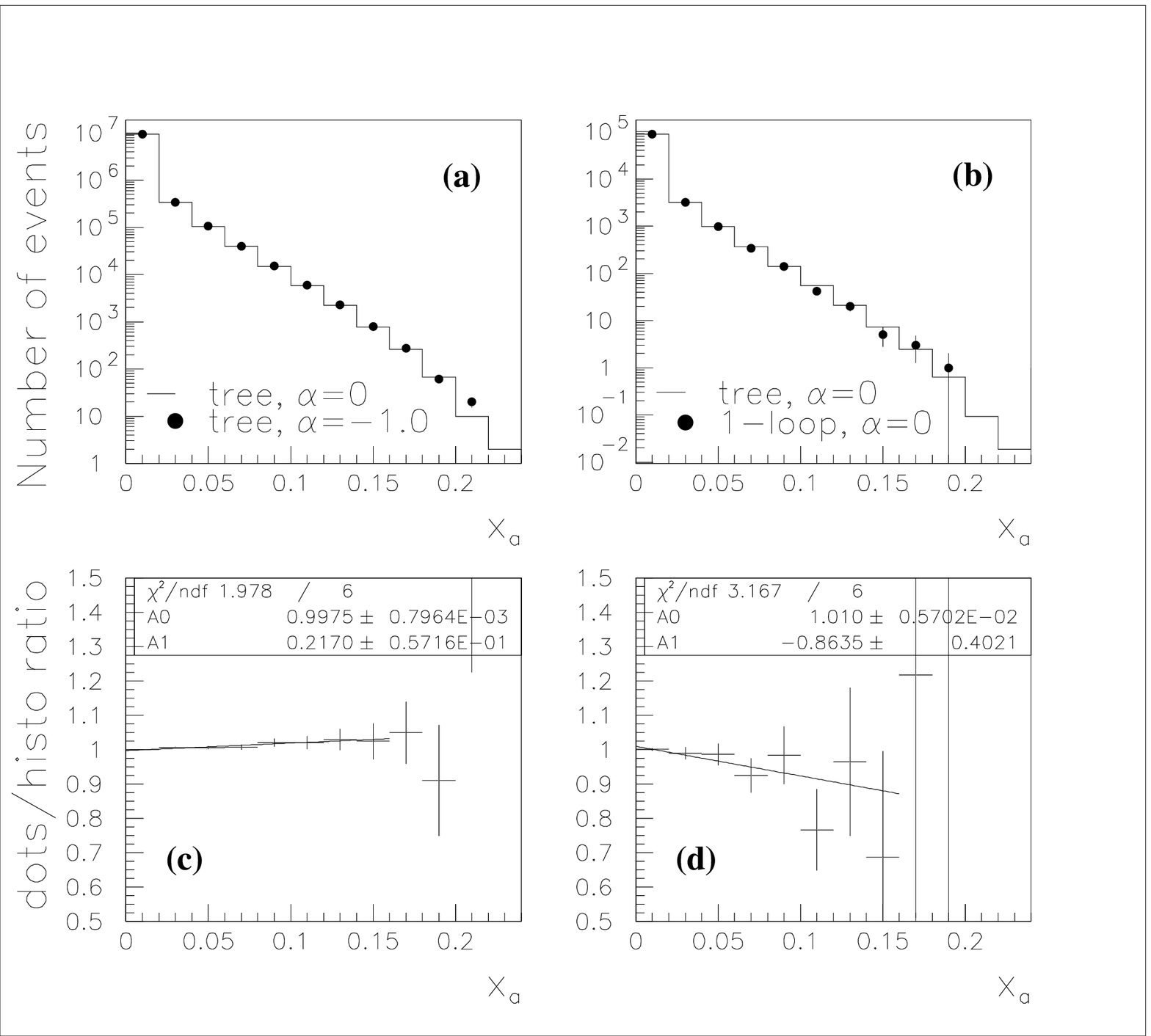}
\caption{\label{fig:xa_pee}
         (a) $x_a$ in $\pzees$ events using the tree--level 
         differential rate with $\alpha=0$ (dots) and with $\alpha=-1.0$ 
         (histogram). 
         (b) $x_a$ in $\pzees$ events using the tree--level 
         differential rate with $\alpha=0$ (dots) and the corrected rate 
         for events with $x_{4e} > 0.95$, also with $\alpha=0$ 
         (histogram).
         The ratio of the dots to the histogram in both cases are shown
         in (c) and (d).} 
\end{figure}

\begin{figure}
\includegraphics[width=3.4in]{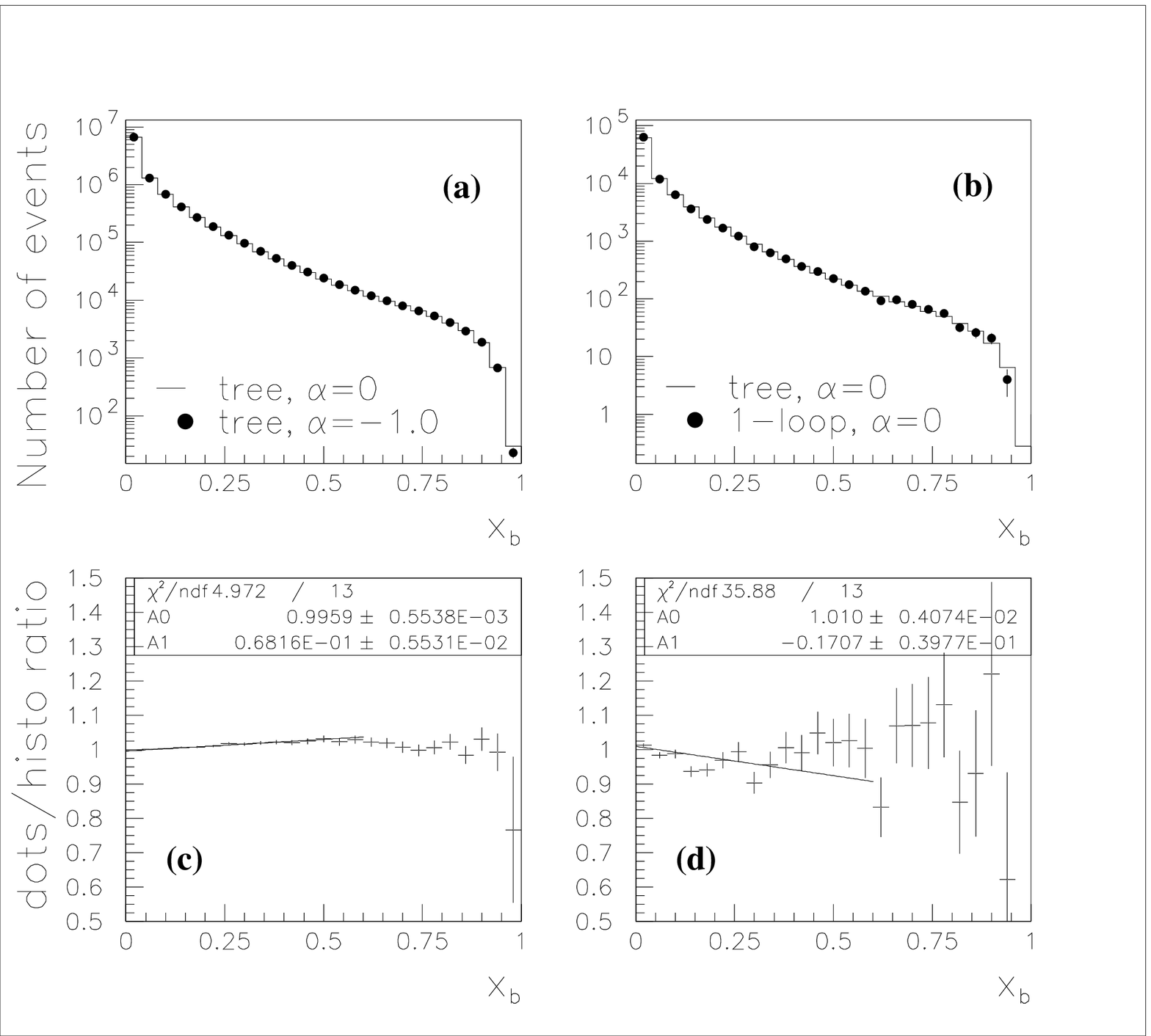}
\caption{\label{fig:xb_pee}
         (a) $x_b$ in $\pzees$ events using the tree--level 
         differential rate with $\alpha=0$ (dots) and with $\alpha=-1.0$ 
         (histogram). 
         (b) $x_b$ in $\pzees$ events using the tree--level 
         differential rate with $\alpha=0$ (dots) and the corrected rate 
         for events with $x_{4e} > 0.95$, also with $\alpha=0$ 
         (histogram).
         The ratio of the dots to the histogram in both cases are shown
         in (c) and (d).} 
\end{figure}

The first variable that we will look at is $x$ which is modified by both the
existence of a form factor and the inclusion of the radiative corrections.
In all cases we set $\beta=0$ in the DIP form factor model.
Figs.~\ref{fig:xa_pee} and~\ref{fig:xb_pee} show the distribution of $x_a$ 
and $x_b$, respectively, for $\pzee$ events. The plot on the left compares 
the distribution using the tree--level matrix element with no form factor 
($\alpha=0$) to that using the same matrix element but with $\alpha=-1.0$. 
The plot on the right compares the distribution using the tree--level 
matrix element with no form factor to that using the radiatively--corrected 
matrix element also with no form factor.
We have provided a linear fit to the ratio over some reasonable range on a
scale appropriate for comparing the two effects. For
the form factor comparisons, the dependence should be primarily linear. This
is not the case for the radiative corrections in general. The $\chi^2$ per
degree of freedom is included as a measure of the linearity. 
It can be seen that the form factor has a much smaller effect on the $x$
distribution than the radiative corrections do. This is not too
surprising since the range of accessible $q^2$ values for the $\pz$ decay
is relatively small in addition to being far from our assumed $\rho$ pole.

\begin{figure}
\includegraphics[width=3.4in]{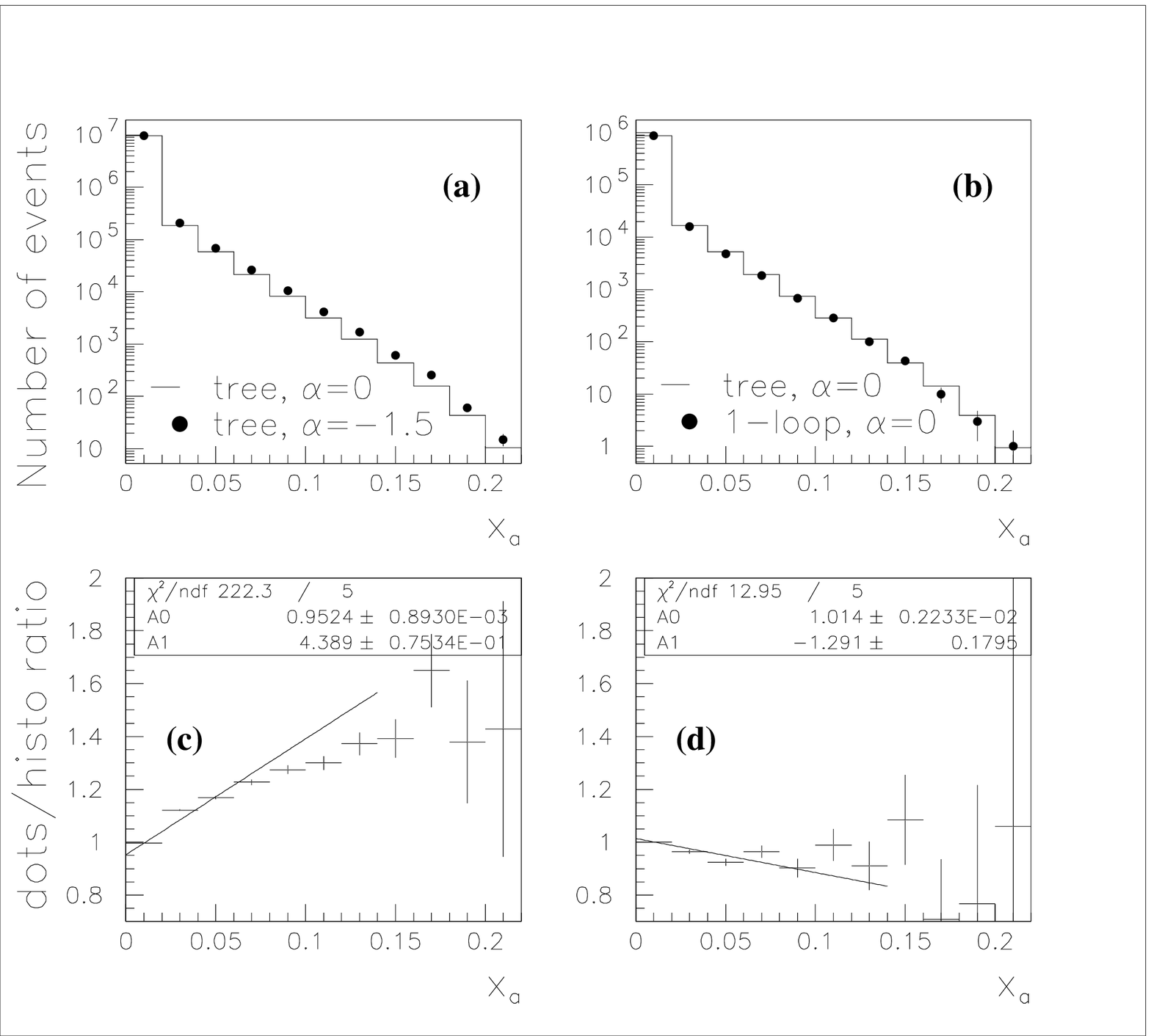}
\caption{\label{fig:xa_kee}
         (a) $x_a$ in $\klees$ events using the tree--level 
         differential rate with $\alpha=0$ (dots) and with $\alpha=-1.5$ 
         (histogram). 
         (b) $x_a$ in $\klees$ events using the tree--level 
         differential rate with $\alpha=0$ (dots) and the corrected rate 
         for events with $x_{4e} > 0.95$, also with $\alpha=0$ 
         (histogram).
         The ratio of the dots to the histogram in both cases are shown
         in (c) and (d).} 
\end{figure}

\begin{figure}
\includegraphics[width=3.4in]{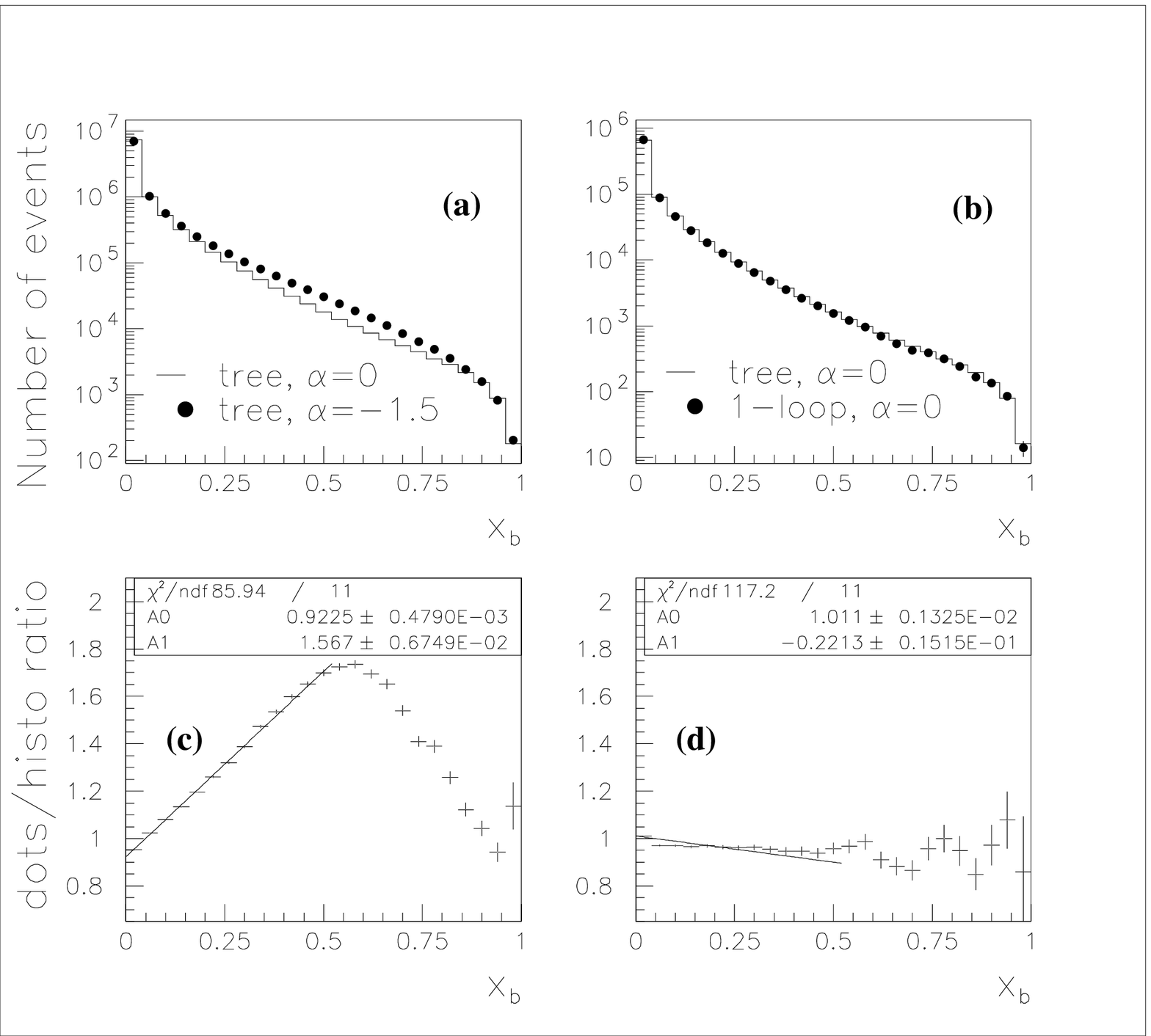}
\caption{\label{fig:xb_kee}
         (a) $x_b$ in $\klees$ events using the tree--level 
         differential rate with $\alpha=0$ (dots) and with $\alpha=-1.5$ 
         (histogram). 
         (b) $x_b$ in $\klees$ events using the tree--level 
         differential rate with $\alpha=0$ (dots) and the corrected rate 
         for events with $x_{4e} > 0.95$, also with $\alpha=0$ 
         (histogram).
         The ratio of the dots to the histogram in both cases are shown
         in (c) and (d).} 
\end{figure}

\begin{figure}
\includegraphics[width=3.4in]{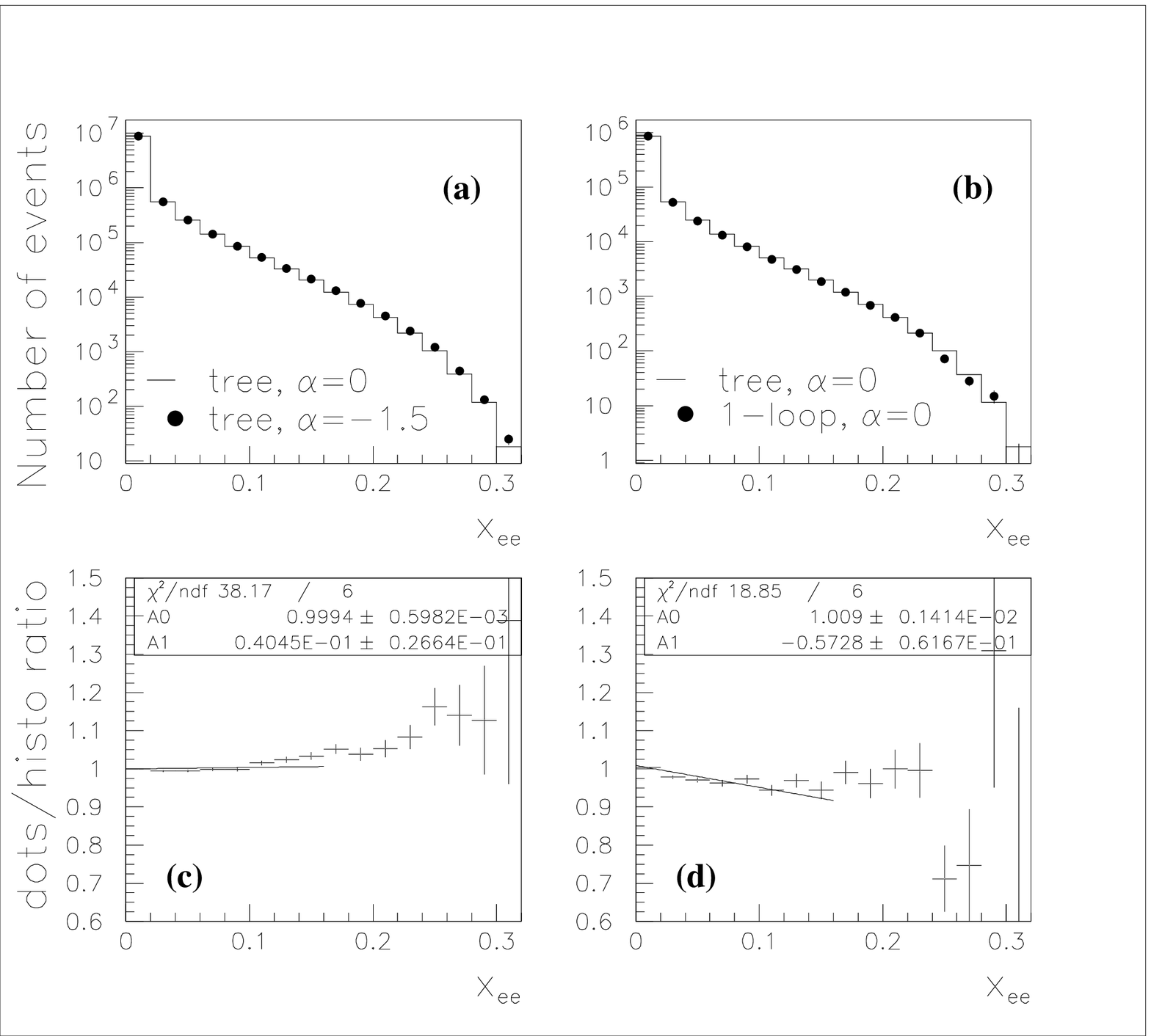}
\caption{\label{fig:xa_kem}
         (a) $x_{ee}$ in $\klems$ events using the tree--level 
         differential rate with $\alpha=0$ (dots) and with $\alpha=-1.5$ 
         (histogram). 
         (b) $x_{ee}$ in $\klems$ events using the tree--level 
         differential rate with $\alpha=0$ (dots) and the corrected rate 
         for events with $x_{4e} > 0.95$, also with $\alpha=0$ 
         (histogram).
         The ratio of the dots to the histogram in both cases are shown
         in (c) and (d).} 
\end{figure}

\begin{figure}
\includegraphics[width=3.4in]{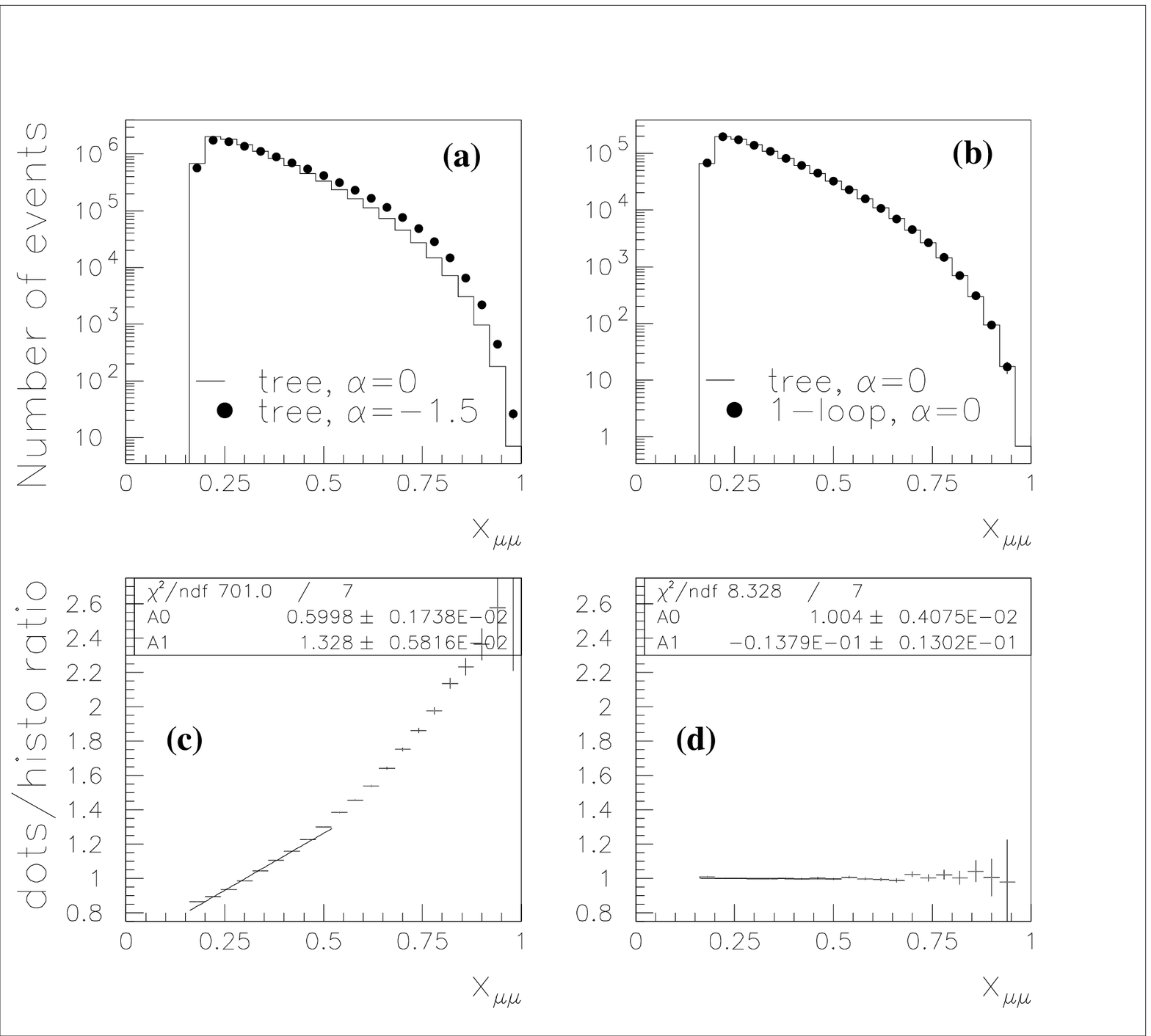}
\caption{\label{fig:xb_kem}
         (a) $x_{\mu\mu}$ in $\klems$ events using the tree--level 
         differential rate with $\alpha=0$ (dots) and with $\alpha=-1.5$ 
         (histogram). 
         (b) $x_{\mu\mu}$ in $\klems$ events using the tree--level 
         differential rate with $\alpha=0$ (dots) and the corrected rate 
         for events with $x_{4e} > 0.95$, also with $\alpha=0$ 
         (histogram).
         The ratio of the dots to the histogram in both cases are shown
         in (c) and (d).} 
\end{figure}

For the kaon modes, we observe that the form factor has a much larger effect
on the $x$ distribution than the radiative corrections do.
Figs.~\ref{fig:xa_kee} and~\ref{fig:xb_kee} show the distribution of $x_a$ 
and $x_b$, respectively, for $\klee$ events. The plot on the left compares 
the distribution using the tree--level matrix element with no form factor 
($\alpha=0$) to that using the same matrix element but with $\alpha=-1.5$. 
The plot on the right compares the distribution using the tree--level 
matrix element with no form factor to that using the radiatively--corrected 
matrix element also with no form factor.
The roll off at high $x$ in plot (c) of Fig.~\ref{fig:xb_kee} is due to
presence of the exchange diagram in this mode. Figs.~\ref{fig:xa_kem} 
and~\ref{fig:xb_kem} show the same distributions for $\klem$ events. 
Here there are no pairing ambiguities and we plot $x_{ee}$ and $x_{\mu\mu}$.
It can be seen that there is no roll off in plot (c) of
Fig.~\ref{fig:xb_kem}, and furthermore, a small quadratic dependence is
observable. While the $x$ of the $ee$ pair is slightly modified by the
radiative corrections, the $x$ of the $\mu\mu$ pair does not change at
all. This is as expected for the massive muons.

\begin{figure}
\includegraphics[width=3.4in]{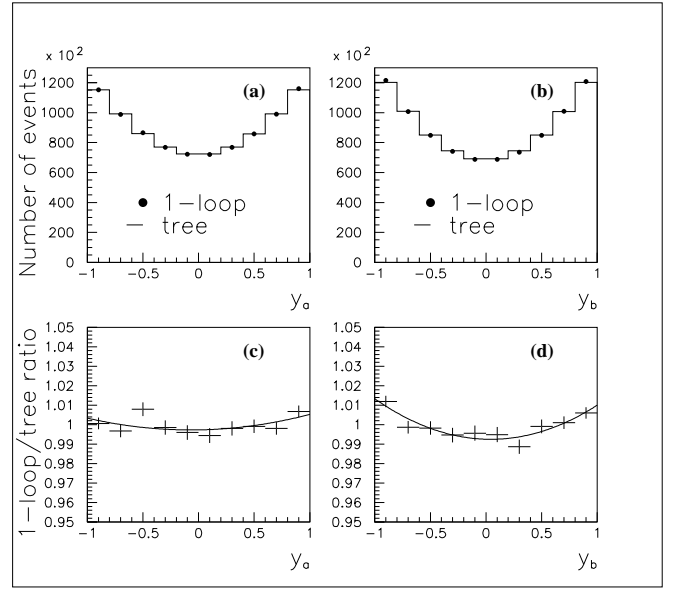}
\caption{\label{fig:yab_kee}(a) $y_a$ in $\klees$ events using the tree--level 
         differential rate (dots) and the corrected rate for events with 
         $x_{4e} > 0.95$ (histogram). (b) $y_b$ for the same events. 
         The ratio of the corrected distributions to the tree--level 
         distributions are shown in (c) and (d).} 
\end{figure}

Fig.~\ref{fig:yab_kee} shows the distribution of $y_a$ and $y_b$ for 
the tree level differential rate and the radiatively corrected differential 
rate for $\klee$ events. 
The effect here is quite small. Since $y$ is a measure of the energy 
asymmetry of the lepton pair, it is seen that the radiative corrections tend
to make the pairs slightly more asymmetric on average.

\begin{figure}
\includegraphics[width=3.4in]{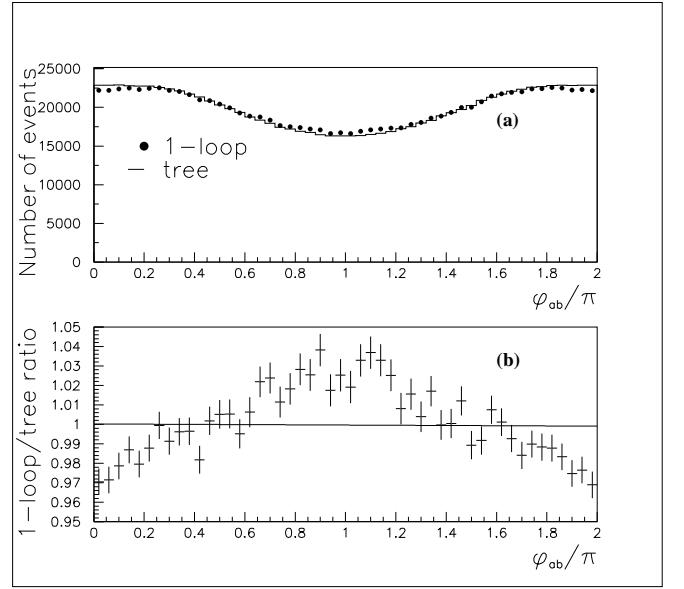}
\caption{\label{fig:phikmm}(a) The distribution of $\phi_{ab}$ in $\klmms$ 
         events 
         using the tree--level differential rate (dots) and the corrected 
         rate for events with $x_{4e} > 0.95$(histogram). 
         (b) The ratio of the corrected distribution to the tree--level 
         distribution.} 
\end{figure}

The effect on the $\phi$ distribution is due entirely to the 5--point
diagram. Fig.~\ref{fig:phikmm} shows a comparison of the distribution of 
$\phi_{ab}$ generated with the tree--level matrix element to the same 
distribution generated with the radiative corrections, for $\klmm$ events.
The enhancement at $\phi_{ab}=\pi$ and the corresponding depletion at
$\phi_{ab}=0=2\pi$ can be understood in terms of Coulomb interaction 
between the final state particles. The configuration at $\phi=0$ has all 
leptons in a plane with the same sign particles near each other. 
The effect is only observable in the $\klmm$ decay where the leptons in 
each pair are usually well separated. 

\section{Conclusions}
The main conclusion that can be drawn from these distributions is that the
radiative corrections are extremely important for extracting a form factor
in the $\pzee$ mode. For the kaon modes, the form factor has a larger impact
on the $x$ distribution and the modification of the distribution due to the
radiative corrections is less important. The only published result
for the $\klee$ mode~\cite{ktev_kleeee} quotes 
$\alpha_{DIP} = -1.1 \pm 0.6 (\text{stat})$. Fig.~\ref{fig:xb_kee} shows
that at most the radiative corrections would change the slope by $0.22$,
which while significant, is smaller than the current experimental error. 
Likewise, for the $\klem$ mode, the latest result\cite{ktev_klem2} based 
on the invariant mass shape is $\alpha_{DIP} = -4.53^{+1.81}_{-2.70}$. The 
present experimental error is again larger than the impact of the radiative 
corrections on the mass distribution.

As for the extraction of the mixing angle $\zeta$ from the observed $\phi$
distribution, the radiative corrections can be safely neglected at present.

The two publications above quote an integrated rate (normalized to the
two--photon rate) of $(6.24 \pm 0.34) \cdot 10^{-5}$ for $\klee$ and 
$(4.51 \pm 0.42) \cdot 10^{-6}$ for $\klem$, where the errors are purely
statistical. These results are in good agreement with our predictions when
both the radiative corrections and a form factor with $\alpha=-1.5$ are
included. For $\klee$, the two effects offset and the net result is an
increase of just less than $2\%$ over the tree--level rate with no form
factor. In $\klem$, the form factor is the dominant effect. 

\begin{acknowledgments}
This work was supported by Department of Energy grant DE-FG03-95ER40894 
and by NSF/REU grant PHY-0097381. The authors also acknowledge useful 
comments from Prof. L. M. Sehgal.
\end{acknowledgments}

\appendix
\section{Kinematics\label{sec:kine}}
The four particle final state can be kinematically described by considering
subsystems containing only two particles. Consider the system composed of
two particles with momenta $p_i$ and $p_j$ and total momentum 
$p_{ij} = p_i + p_j$ and mass squared $m_{ij}^2 = p_{ij}^2$. We will define 
a dimensionless dot product of any two vectors $p_i$ and $p_j$ as
\begin{equation}
z_{ij} = 2 (p_i \cdot p_j)/p_{ij}^2 = 1 - x_i - x_j,
\end{equation}
where
\begin{gather}
x_i = p_i^2/p_{ij}^2, \quad 
x_j = p_j^2/p_{ij}^2.
\end{gather}
The energy and momentum of each particle in the two--particle CM frame are
\begin{gather}
E_i^* = m_{ij}(1 + \delta_{ij})/2, \quad
E_j^* = m_{ij}(1 - \delta_{ij})/2, \\
p^*   = m_{ij}\lambda_{ij}/2,
\end{gather}
where
\begin{align}
\delta_{ij}  & = x_i - x_j, \\
\lambda_{ij} & = \sqrt{z_{ij}^2 - w_{ij}^2},\\
w_{ij}       & = 2 \sqrt{x_i x_j}.
\end{align}
Occasionally we will use symbols like $z_{i,jk}$ whose meaning is
interpreted as $z(p_i, p_j+p_k)$.

Now consider a three body system composed of momenta $p_i$, $p_j$, and
$p_k$. There are two phase space variables needed to describe the system. The
first one will be $x_{ij} = p_{ij}^2/p_{ijk}^2$. The other one is defined 
in the $ij$ CM frame as the cosine of the angle between the direction 
of particle $i$ and particle $k$,
\begin{equation}
\cos{\theta_{ij,k}} = \frac{2 p_k \cdot (p_i - p_j) 
                              - p_{ijk}^2 \delta_{ij} z_{ij,k}}
                           {p_{ijk}^2 \lambda_{ij} \lambda_{ij,k}},
\end{equation}
A more convenient variable that will be used in place of 
$\cos{\theta_{ij,k}}$ is
\begin{equation}
y_{ij} = \lambda_{ij} \cos{\theta_{ij,k}},
\end{equation}
where $k$ will always refer to the total momentum minus the $ij$ momentum.

Finally, the four body final state requires five phase space variables to
uniquely describe it. We will use the $x$ and $y$ values for the two lepton
pairs plus the angle between the normals of the planes defined by each pair
in the overall CM frame. The first four variables are
\begin{align}
x_{12} &= p_{12}^2/M^2,\\
x_{34} &= p_{34}^2/M^2,\\
y_{12} &= \frac{2 p_{34} \cdot (p_1 - p_2)}{M^2 \lambda},\\
y_{34} &= \frac{2 p_{12} \cdot (p_3 - p_4)}{M^2 \lambda},
\end{align}
where the second term in the numerator of the $y$'s vanishes since
$\delta_{12} = \delta_{34} = 0$. Any use of $z$, $w$, or $\lambda$ without
subscripts will refer to the functions of $x_{12}$ and $x_{34}$, so 
$\lambda = \lambda_{12,34}$ for instance.
The last phase space variable is defined as
\begin{equation}
\phi = \tan^{-1}{(\sin{\phi}/\cos{\phi})},
\end{equation}
where
\begin{align}
\sin{\phi} & = 
  \frac{16 \epsilon_{\mu\nu\rho\sigma} p_1^\mu p_2^\nu p_3^\rho p_4^\sigma}
       {M^4 \lambda w 
          \sqrt{(\lambda_{12}^2 - y_{12}^2) (\lambda_{34}^2 - y_{34}^2)}},\\
\cos{\phi} & = 
  \frac{M^2 z y_{12} y_{34} - 2 (p_1 - p_2) \cdot (p_3 - p_4)}
       {M^2 w
          \sqrt{(\lambda_{12}^2 - y_{12}^2) (\lambda_{34}^2 - y_{34}^2)}}.
\end{align}
The angle $\phi$ is defined so that at $\phi=0$ the two pairs lie in a plane
with the like--sign particles adjacent to each other. The orientation of
$\phi=\pi$ again has both pairs in a plane, but with the opposite signed 
particles adjacent.

The general expression for the phase space integral is
\begin{equation}
d^8 \Phi = \frac{1}{(2\pi)^8} 
           \frac{d^3 p_1 d^3 p_2 d^3 p_3 d^3 p_4}
                {16 E_{p_1} E_{p_2} E_{p_3} E_{p_4}}
           \delta^4(P - p_1 - p_2 - p_3 - p_4).
\end{equation}
Upon integrating out the $\delta$--functions, integrating over the Euler
angles, and changing variables to those listed above, the phase space 
reduces to
\begin{equation}
d^5 \Phi = \mathcal{S} \frac{M^4}{2^{14}\pi^6} \lambda 
           d x_{12} d x_{34} d y_{12} d y_{34} d \phi,
\end{equation}
where the factor $\mathcal{S}$ is a symmetry factor which is required for 
modes containing identical particles in the final state. The double Dalitz
modes with identical particles contain two sets, thus requiring two factors
of $1/2$. So $\mathcal{S} = 1/4$ if the final state contains identical 
particles, and $\mathcal{S} = 1$ otherwise.

When there are identical leptons in the final state, the amplitudes for the
exchange diagrams have the same algebraic form as for the non--exchange
diagrams except that the kinematic variables $x_{12}$, $x_{34}$, $y_{12}$,
$y_{34}$, and $\phi$ are replaced by $x_{14}$, $x_{23}$, $y_{14}$, $y_{23}$,
and $\phi_{14,23}$. These exchange variables will in general be
functions of all five of the non--exchange variables. As is turns out, we
only need explicit representations for $x_{14}$ and $x_{23}$. These are
given by
\begin{align}
x_{14} &= [1 - \lambda_{12}^2 x_{12} - \lambda_{34}^2 x_{34} 
           - \lambda (y_{12}-y_{34}) - z y_{12} y_{34}\nn\\
       &\quad\quad
           + w \sqrt{(\lambda_{12}^2 - y_{12}^2)(\lambda_{34}^2 - y_{34}^2)} 
             \cos{\phi}]/4,\\
x_{23} &= [1 - \lambda_{12}^2 x_{12} - \lambda_{34}^2 x_{34} 
           + \lambda (y_{12}-y_{34}) - z y_{12} y_{34}\nn\\
       &\quad\quad
           + w \sqrt{(\lambda_{12}^2 - y_{12}^2)(\lambda_{34}^2 - y_{34}^2)} 
             \cos{\phi}]/4.
\end{align}

\section{Meson--$\g\g$ Couplings\label{sec:couple}}
In this section we will work out the explicit form of the two--photon
couplings, allowing for photons of arbitrary mass, using the polarization 
vectors in the helicity basis. The general form of the coupling is 
\begin{equation}
\mathcal{H}_{\lambda_1 \lambda_2} = H_{\mu\nu\rho\sigma} 
    k_1^\mu \epsilon_{\lambda_1}^{* \nu} 
    k_2^\rho \epsilon_{\lambda_2}^{* \sigma},
\end{equation}
where $H$ is either
\begin{subequations} \label{eq:coupling}
\begin{align}
H_{\mu\nu\rho\sigma}^P & = \frac{2}{M} 
   \mathcal{F}_P\, \epsilon_{\mu\nu\rho\sigma},\\
H_{\mu\nu\rho\sigma}^S & = \frac{2}{M} 
   \mathcal{F}_S (  g_{\mu\rho} g_{\nu\sigma} 
                  - g_{\mu\sigma} g_{\nu\rho}).
\end{align}
\end{subequations}
The three polarization vectors for a massive photon in the helicity basis are
chosen to be
\begin{subequations}
\begin{align}
\epsilon^+(\pm \vhat{z}) & = \left(0, 1, \pm i, 0\right)/\sqrt{2},\\
\epsilon^-(\pm \vhat{z}) & = \left(0, 1, \mp i, 0\right)/\sqrt{2},\\
\epsilon^0(\pm \vhat{z}) & = \left(k, 0, 0, \pm E\right)/\sqrt{k^2},
\end{align}
\end{subequations}
for a photon traveling in the $\pm \vhat{z}$ direction. With these 
polarization vectors, one finds three couplings for the scalar case
\begin{equation} \label{eq:scouple}
\mathcal{H}_{\lambda_1 \lambda_2}^S = 
  \begin{cases}
    - M \mathcal{F}_S z &, \lambda_1 = \lambda_2 = +,\\
    - M \mathcal{F}_S z &, \lambda_1 = \lambda_2 = -,\\
    + M \mathcal{F}_S w &, \lambda_1 = \lambda_2 = 0,
  \end{cases}
\end{equation}
where $z$ and $w$ are defined in 
Appendix~\ref{sec:kine}. The longitudinal contribution vanishes for the 
pseudoscalar case, and one finds only two couplings
\begin{equation} \label{eq:pcouple}
\mathcal{H}_{\lambda_1 \lambda_2}^P = 
  \begin{cases}
    +i M \mathcal{F}_P \lambda &, \lambda_1 = \lambda_2 = +,\\
    +i M \mathcal{F}_P \lambda &, \lambda_1 = \lambda_2 = -,\\
                             0 &, \lambda_1 = \lambda_2 = 0,
  \end{cases}
\end{equation}
where $\lambda$ is also defined in Appendix~\ref{sec:kine}. There are three
interesting differences between the scalar and the pseudoscalar couplings.
First, assuming that $\delta=0$, there is a relative phase between them.
Additionally, the transverse couplings differ in the kinematic factor. And
lastly, there is the additional scalar coupling due to the contribution 
from longitudinally polarized photons. Where as the transverse couplings 
go like $\lambda$ or $z$, both of which are $\mathcal{O}(1)$ on average, 
the longitudinal coupling goes like $w$, which is $\mathcal{O}(x)$, making 
its contribution less significant.

\section{Double Dalitz Interference\label{sec:interf}}
The interference between the tree--level direct and exchange contributions
for modes with identical leptons is a sum of three terms
\begin{align}
2\real&{(\mel_1^* \mel_2)} = \frac{2^3 \pi^2 \alpha^2 \tilde{g}^2}
                                 {M^2 x_{12} x_{34} x_{14} x_{23}}\nn\\
& \times \bigg\{A f_P(x_{12},x_{34}) f_P(x_{14},x_{23}) \cos^2\zeta\nn\\ 
& \quad       + B [  f_P(x_{12},x_{34}) f_S(x_{14},x_{23})\nn\\ 
& \quad\quad\quad  + f_P(x_{14},x_{23}) f_S(x_{12},x_{34}) ] 
                  \sin\zeta \cos\zeta\nn\\
& \quad       + C f_S(x_{12},x_{34}) f_S(x_{14},x_{23}) \sin^2\zeta\bigg\},
\end{align}
where
\begin{align}
A &= \lambda^2\{
         2 \Xi^2 + 8 \eta^4
       + \Xi \left[8 \eta^2 - z (y_{12}+y_{34})^2\right]\nn\\
  & \quad\quad\quad   - w^2 (1+y_{12}y_{34})(2-y_{12}^2-y_{34}^2)\nn\\
  & \quad\quad\quad   + 4 \eta^2 (y_{12}+y_{34})(x_{12}y_{12}+x_{34}y_{34})
      \},
\end{align}
\begin{align}
B &= -2 \lambda
      [\eta^2 (x_{12}+x_{34}-x_{14}-x_{23})\nn\\
  & \quad\quad\quad + 4(x_{12}x_{34} - x_{14}x_{23})] \Xi \tan\phi,
\end{align}
\begin{align}
C &= -\Xi^3 z - \Xi^2 [6 \eta^2 z - w^2 - (2z^2+w^2) y_{12} y_{34}]\nn\\
  & - \Xi [z (z^2-2w^2) + z (z^2+2w^2) y_{12}^2 y_{34}^2
            + 2 z^3 y_{12} y_{34}\nn\\
  & \quad\quad   - 2 \eta^2 (z^2 + w^2) (1 + 3 y_{12} y_{34})
            + 8 \eta^4 z]\nn\\
  & + w^2 z^2 y_{12}^3 y_{34}^3 - w^2 z^2 y_{12} y_{34}
    + w^2 (3z^2 - 2w^2) y_{12}^2 y_{34}^2\nn\\
  & - 2 w^2 (z^2 - w^2)(y_{12}^2 + y_{34}^2) + w^2 (z^2 - 2w^2)\nn\\
  & - 2\eta^2[3 w^2 z y_{12}^2 y_{34}^2
    - 2 (z^2 - z^3 - w^2) y_{12} y_{34}\nn\\
  & \quad\quad\quad + 2 z (z^2 - 3w^2/2) - 2(z^2 + w^2)]\nn\\
  & - 8 \eta^4 [(z-z^2) - w^2 y_{12}y_{34}],
\end{align}
where 
$\Xi = w \sqrt{(\lambda_{12}^2-y_{12}^2)(\lambda_{34}^2-y_{34}^2)} \cos\phi$
and $\eta^2 = 4 m^2/M^2$.
The exchange variables $x_{14}$ and $x_{23}$ are defined in
Appendix~\ref{sec:kine} in terms of the five non--exchange phase space
variables. The term proportional to $\cos^2\zeta$ results from interference
between a pseudoscalar coupling in both the direct and exchange graphs,
while the one proportional to $\sin^2\zeta$ is due to scalar couplings in
both graphs, and the one proportional to $\sin\zeta \cos\zeta$ is due to a 
pseudoscalar coupling in one graph and a scalar coupling in the other.

\section{5--Point Function\label{sec:hf}}
The matrix element for the 5--point diagram is composed of tensor integrals
with five propagators in the denominator. One can express tensor integrals
in terms of lower rank tensor integrals with the same number of propagators 
and lower rank tensors with fewer propagators~\cite{oldenborgh}. In the end, 
every tensor integral can be decomposed into scalar 2--, 3--, 4--, and 
5--point functions. The scalar 5--point function is not independent and can 
itself be expressed in terms of scalar 4--point functions. 

This appendix will outline our procedure for first reducing the tensor 
integrals to scalar integrals, and then computing the scalar integrals in
closed form.

\begin{widetext}
\subsubsection{Tensor Integrals}%
Begin by defining
\begin{align}
I_{50}(k_1,k_2,k_3,k_4)       & = 
  \int \frac{d^4 t}{(2\pi)^4} \frac{1}{N_1 N_2 N_3 N_4 N_5}, &
I_{52}^{\mu\nu}(k_1,k_2,k_3,k_4) & = 
  \int \frac{d^4 t}{(2\pi)^4} \frac{t^\mu t^\nu}{N_1 N_2 N_3 N_4 N_5}, \\
I_{51}^{\mu}(k_1,k_2,k_3,k_4) & = 
  \int \frac{d^4 t}{(2\pi)^4} \frac{t^\mu}{N_1 N_2 N_3 N_4 N_5}, &
I_{53}^{\mu\nu\rho}(k_1,k_2,k_3,k_4) & = 
  \int \frac{d^4 t}{(2\pi)^4} \frac{t^\mu t^\nu t^\rho}{N_1 N_2 N_3 N_4 N_5},
\end{align}
where
\begin{gather}
N_1 = t^2 - \mu_1^2, \quad \quad
N_2 = (t+k_1)^2 - \mu_2^2, \quad \quad
N_3 = (t+k_1+k_2)^2 - \mu_3^2, \\
N_4 = (t+k_1+k_2+k_3)^2 - \mu_4^2, \quad \quad
N_5 = (t+k_1+k_2+k_3+k_4)^2 - \mu_5^2,
\end{gather}
where $\mu_i$ is an internal mass and the $k_i$ are external momenta.

The original reduction scheme of Ref.~\cite{passarino}, while theoretically
sound, suffers from uncontrollable numerical inaccuracies. To avoid this
problem, we follow the procedure suggested in Ref.~\cite{oldenborgh}, and
use a reduction scheme based on the Schouten identity which utilizes Gram 
determinants to express any tensor integral as a sum of integrals, one 
with the same number of propagators and the rest with one less propagator, 
and all with the power of the loop momentum reduced by one. The identity 
has the following form
\begin{align}
t^\mu \epsilon^{k_1 k_2 k_3 k_4} 
 & =  (t \cdot k_1) \epsilon^{\mu k_2 k_3 k_4}
    + (t \cdot k_2) \epsilon^{k_1 \mu k_3 k_4}
    + (t \cdot k_3) \epsilon^{k_1 k_2 \mu k_4}
    + (t \cdot k_4) \epsilon^{k_1 k_2 k_3 \mu},\\
 & = v^\mu - \frac{1}{2} \left(  N_1 \epsilon^{\mu k_2 k_3 k_4}
                               - N_2 \epsilon^{\mu (k_1+k_2) k_3 k_4}
                               + N_3 \epsilon^{\mu k_1 (k_2+k_3) k_4}
                               - N_4 \epsilon^{\mu k_1 k_2 (k_3+k_4)}
                               + N_5 \epsilon^{\mu k_1 k_2 k_3}\right),
\end{align}
where
\begin{equation} \label{eq:vvec}
v^\mu =  (s \cdot k_1) \epsilon^{\mu k_2 k_3 k_4}
       - (s \cdot k_2) \epsilon^{\mu k_1 k_3 k_4}
       + (s \cdot k_3) \epsilon^{\mu k_1 k_2 k_4}
       - (s \cdot k_4) \epsilon^{\mu k_1 k_2 k_3},
\end{equation}
and $s^\mu$ is defined in terms of its dot products
\begin{align}
s \cdot k_1 & = \frac{1}{2}\left[\mu_2^2 - \mu_1^2 - k_1^2\right], &
s \cdot k_3 & = \frac{1}{2}\left[\mu_4^2 - \mu_1^2 - (k_1+k_2+k_3)^2\right] 
     - s \cdot k_1 - s \cdot k_2,\\
s \cdot k_2 & = \frac{1}{2}\left[\mu_3^2 - \mu_1^2 - (k_1+k_2)^2\right] 
     - s \cdot k_1, &
s \cdot k_4 & = \frac{1}{2}\left[\mu_5^2 - \mu_1^2 -(k_1+k_2+k_3+k_4)^2\right] 
     - s \cdot k_1 - s \cdot k_2 - s \cdot k_3.
\end{align}
The notation $\epsilon^{k_1 k_2 k_3 k_4}$ is shorthand for 
$\epsilon^{\mu\nu\rho\sigma} k_{1\mu} k_{2\nu} k_{3\rho} k_{4\sigma}$.

The reduction then proceeds as follows:
\begin{align}
I_{51}^\mu & = \frac{1}{\epsilon^{k_1 k_2 k_3 k_4}} \bigg\{
 v^\mu I_{50} - \frac{1}{2} \bigg[
        \epsilon^{\mu k_2 k_3 k_4}       I_{40}^{(1)}
      - \epsilon^{\mu (k_1+k_2) k_3 k_4} I_{40}^{(2)} 
      + \epsilon^{\mu k_1 (k_2+k_3) k_4} I_{40}^{(3)}
      - \epsilon^{\mu k_1 k_2 (k_3+k_4)} I_{40}^{(4)}\nn\\ 
 & \quad \quad \quad \quad \quad \quad \quad \quad \quad \quad \quad
      + \epsilon^{\mu k_1 k_2 k_3}       I_{40}^{(5)}  \bigg] \bigg\},\\
I_{52}^{\mu\nu} & = \frac{1}{\epsilon^{k_1 k_2 k_3 k_4}} \bigg\{
 v^\mu I_{51}^\nu - \frac{1}{2} \bigg[
        \epsilon^{\mu k_2 k_3 k_4} 
          \bigg(I_{41}^{(1)\nu} - k_1^\nu I_{40}^{(1)}\bigg)
      - \epsilon^{\mu (k_1+k_2) k_3 k_4} I_{41}^{(2)\nu} 
      + \epsilon^{\mu k_1 (k_2+k_3) k_4} I_{41}^{(3)\nu}\nn\\
 & \quad \quad \quad \quad \quad \quad \quad \quad \quad \quad \quad
      - \epsilon^{\mu k_1 k_2 (k_3+k_4)} I_{41}^{(4)\nu} 
      + \epsilon^{\mu k_1 k_2 k_3}       I_{41}^{(5)\nu}  \bigg] \bigg\},\\
I_{53}^{\mu\nu\rho} & = \frac{1}{\epsilon^{k_1 k_2 k_3 k_4}} \bigg\{
 v^\mu I_{52}^{\nu\rho}
   - \frac{1}{2} \bigg[ \epsilon^{\mu k_2 k_3 k_4}   \bigg(
             I_{42}^{(1)\nu\rho} - k_1^\nu I_{41}^{(1)\rho} 
           - k_1^\rho I_{41}^{(1)\nu} + k_1^\nu k_1^\rho I_{40}^{(1)} \bigg)
      - \epsilon^{\mu (k_1+k_2) k_3 k_4} I_{42}^{(2)\nu\rho}\nn\\ 
 & \quad \quad \quad \quad \quad \quad \quad \quad \quad \quad \quad
      + \epsilon^{\mu k_1 (k_2+k_3) k_4} I_{42}^{(3)\nu\rho}
      - \epsilon^{\mu k_1 k_2 (k_3+k_4)} I_{42}^{(4)\nu\rho} 
      + \epsilon^{\mu k_1 k_2 k_3}       I_{42}^{(5)\nu\rho}  \bigg] \bigg\},
\end{align}
where for any 
$I_{40} = \int \frac{d^4 t}{(2\pi)^4} \frac{1}{N_1 N_2 N_3 N_4}$,
\begin{align}
I_{41}^{\mu} & = \frac{1}{2 \delta^{k_1 k_2 k_3}_{k_1 k_2 k_3}} \bigg\{
 \delta^{s \alpha \beta}_{k_1 k_2 k_3} \delta^{\mu \alpha \beta}_{k_1 k_2 k_3}
 I_{40} - \bigg[
        \delta^{\mu k_2 k_3}_{k_1 k_2 k_3}       I_{30}^{(1)}
      - \delta^{\mu (k_1+k_2) k_3}_{k_1 k_2 k_3} I_{30}^{(2)}
      + \delta^{\mu k_1 (k_2+k_3)}_{k_1 k_2 k_3} I_{30}^{(3)}
      - \delta^{\mu k_1 k_2}_{k_1 k_2 k_3}       I_{30}^{(4)} \bigg] \bigg\},\\
I_{42}^{\mu\nu} & = \frac{1}{2 \delta^{k_1 k_2 k_3}_{k_1 k_2 k_3}} \bigg\{
 \delta^{s \alpha \beta}_{k_1 k_2 k_3} \delta^{\mu \alpha \beta}_{k_1 k_2 k_3}
 I_{41}^{\nu} - \bigg[
        \delta^{\mu k_2 k_3}_{k_1 k_2 k_3} 
            \bigg(I_{31}^{\nu} - k_1 ^\nu I_{30}^{(1)}\bigg)
      - \delta^{\mu (k_1+k_2) k_3}_{k_1 k_2 k_3} I_{31}^{(2)\nu}\nn\\
 & \quad \quad \quad \quad \quad \quad \quad \quad \quad \quad \quad
   \quad \quad \quad \quad
      + \delta^{\mu k_1 (k_2+k_3)}_{k_1 k_2 k_3} I_{31}^{(3)\nu}
      - \delta^{\mu k_1 k_2}_{k_1 k_2 k_3}       I_{31}^{(4)\nu} \bigg]\nn\\
 & \quad \quad \quad \quad \quad \quad
 + \frac{2 \epsilon^{k_1 k_2 k_3 \mu} \epsilon^{k_1 k_2 k_3 \nu}}
        {\delta^{k_1 k_2 k_3}_{k_1 k_2 k_3}} \bigg[
      \delta^{s k_1 k_2 k_3}_{s k_1 k_2 k_3} I_{40}
 + \frac{1}{2} \bigg(
      \delta^{(s+k_1) k_2 k_3}_{k_1 k_2 k_3} I_{30}^{(1)}
    - \delta^{s (k_1+k_2) k_3}_{k_1 k_2 k_3} I_{30}^{(2)}\nn\\
 & \quad \quad \quad \quad \quad \quad \quad \quad \quad \quad
   \quad \quad \quad \quad \quad \quad \quad \quad \quad \quad
   \quad \quad \quad
    + \delta^{s k_1 (k_2+k_3)}_{k_1 k_2 k_3} I_{30}^{(3)}
    - \delta^{s k_1 k_2}_{k_1 k_2 k_3}      I_{30}^{(4)} \bigg) \bigg] \bigg\},
\end{align}
and for any $I_{30} = \int \frac{d^4 t}{(2\pi)^4} \frac{1}{N_1 N_2 N_3}$
\begin{equation}
I_{31}^{\mu} = \frac{1}{\delta^{k_1 k_2}_{k_1 k_2}} \bigg\{
  \delta^{s \alpha}_{k_1 k_2} \delta^{\mu \alpha}_{k_1 k_2} I_{30} 
  - \frac{1}{2} \bigg[
        \delta^{\mu k_2}_{k_1 k_2}       I_{20}^{(1)}
      - \delta^{\mu (k_1+k_2)}_{k_1 k_2} I_{20}^{(2)} 
      + \delta^{\mu k_1}_{k_1 k_2}       I_{20}^{(3)} \bigg] \bigg\}.
\end{equation}
The reduction notation has the following meaning: $I_{40}^{(i)}$ is the
4--point function obtained from $I_{50}$ by dropping the $i$th propagator,
$I_{30}^{(j)}$ is the 3--point function obtained from its corresponding
4--point function by dropping the $j$th propagator, and so on.
The Gram determinants that appear in the reduction are kinematic functions
which are defined as
\begin{align}
\delta^{k_1 k_2}_{q_1 q_2} & = 
  \begin{vmatrix}
   q_1 \cdot k_1 & q_1 \cdot k_2\\
   q_2 \cdot k_1 & q_2 \cdot k_2
  \end{vmatrix}, &
\delta^{k_1 k_2 k_3}_{q_1 q_2 q_3} & = 
  \begin{vmatrix}
   q_1 \cdot k_1 & q_1 \cdot k_2 & q_1 \cdot k_3\\
   q_2 \cdot k_1 & q_2 \cdot k_2 & q_2 \cdot k_3\\
   q_3 \cdot k_1 & q_3 \cdot k_2 & q_3 \cdot k_3
  \end{vmatrix}, &
\delta^{k_1 k_2 k_3 k_4}_{q_1 q_2 q_3 q_4} & = 
  \begin{vmatrix}
   q_1 \cdot k_1 & q_1 \cdot k_2 & q_1 \cdot k_3 & q_1 \cdot k_4\\
   q_2 \cdot k_1 & q_2 \cdot k_2 & q_2 \cdot k_3 & q_2 \cdot k_4\\
   q_3 \cdot k_1 & q_3 \cdot k_2 & q_3 \cdot k_3 & q_3 \cdot k_4\\
   q_4 \cdot k_1 & q_4 \cdot k_2 & q_4 \cdot k_3 & q_4 \cdot k_4
  \end{vmatrix}.
\end{align}

The traces that appear in Eq.~\ref{eq:5pmat} can also be reduced 
\begin{subequations}
\label{eq:trace}
\begin{align}
(I_{52})^{\mu}_\mu            &= I_{40}^{(1)},\\
(I_{53})^{\alpha\mu}_\mu      &= I_{41}^{(1)\alpha} 
                                 - k_1^\alpha I_{40}^{(1)},\\
(I_{54})^{\alpha\beta\mu}_\mu &= I_{42}^{(1)\alpha\beta}
                                 - k_1^\alpha I_{41}^{(1)\beta}
                                 - k_1^\beta I_{41}^{(1)\alpha}
                                 + k_1^\alpha k_1^\beta I_{40}^{(1)}.
\end{align}
\end{subequations}

While this procedure is generally much more reliable then that of
Ref.~\cite{passarino}, there are still problems that occur when $v^\mu$ as
defined in Eq.~(\ref{eq:vvec}) is not an independent combination of the
final state momenta. This happens when all the momenta in the parent
particle's rest frame lie in a plane. Even though these configurations form 
a subspace of zero measure in the final state phase space, finite numerical 
precision dictates that they will be generated with non--zero probability 
by the MC program. In this case, we use the identity
\begin{equation}
t^\mu \delta^{k_1 k_2 k_3}_{k_1 k_2 k_3} 
= u^\mu (u \cdot t) + \frac{1}{2} \left[
  \delta^{s \alpha \beta}_{k_1 k_2 k_3} \delta^{\mu \alpha \beta}_{k_1 k_2 k_3}
- N_1 \delta^{\mu k_2 k_3}_{k_1 k_2 k_3}
+ N_2 \delta^{\mu k_3 k_4}_{k_1 k_2 k_3}
- N_3 \delta^{\mu k_4 k_1}_{k_1 k_2 k_3}
+ N_4 \delta^{\mu k_1 k_2}_{k_1 k_2 k_3} \right],
\end{equation}
where $u^\mu = \epsilon^{k_1 k_2 k_3 \mu}$. The first term on the right
vanishes upon integration, allowing $I_{51}^\mu$ to be written as
\begin{equation}
I_{51}^{\mu} = \frac{1}{2 \delta^{k_1 k_2 k_3}_{k_1 k_2 k_3}} \bigg\{
 \delta^{s \alpha \beta}_{k_1 k_2 k_3} \delta^{\mu \alpha \beta}_{k_1 k_2 k_3}
 I_{50} - \bigg[
        \delta^{\mu k_2 k_3}_{k_1 k_2 k_3} I_{40}^{(1)}
      - \delta^{\mu k_3 k_4}_{k_1 k_2 k_3} I_{40}^{(2)}
      + \delta^{\mu k_4 k_1}_{k_1 k_2 k_3} I_{40}^{(3)}
      - \delta^{\mu k_1 k_2}_{k_1 k_2 k_3} I_{40}^{(4)} \bigg] \bigg\}.
\end{equation}
The other tensor integrals can be expanded in a similar manner.
The same problem can arise in the reduction of the 4--point tensor integrals
if the three momenta $k_1$, $k_2$, and $k_3$ are linearly dependent, in
which case this same procedure is reproduced at one lower rank.

In these degenerate cases we have a choice between numerical inaccuracies
resulting from antisymmetric invariants such as $\epsilon^{k_1 k_2 k_3 k_4}$
being very small or inaccuracies resulting from assuming exact linear
dependence. To decide which approximation is better, we do the calculation
of the tensor integrals both ways and check whether identities such as 
those in Eqs.~(\ref{eq:trace})
are satisfied. More than $99\%$ of the time, one of the two methods yields
good agreement for all these ``trace checks''.

\subsubsection{Scalar Integrals}%
The most general 5--point function we will need to consider is
\begin{align}
I_{50} = \mu^{2\epsilon} \int \frac{d^D t}{(2\pi)^D} \bigg\{
 &        \frac{1}{[t^2                   - \mu_1^2 + i\epsilon]
                   [(t+k_1)^2             - \mu_2^2 + i\epsilon]
                   [(t+k_1+k_2)^2         - \mu_3^2 + i\epsilon]}\nn\\ 
 \label{eq:i50}
 & \times \frac{1}{[(t+k_1+k_2+k_3)^2     - \mu_4^2 + i\epsilon]
                   [(t+k_1+k_2+k_3+k_4)^2 - \mu_5^2 + i\epsilon]} \bigg\}.
\end{align}
In the case at hand $\mu_1 = 0$, $\mu_2 = m_1$, $\mu_3 = M_1$, 
$\mu_4 = M_2$, and $\mu_5 = m_2$ where the $m$'s are lepton masses and the
$M$'s are boson masses. In addition, $k_1 = -p_2$, $k_2 = -p_1$, 
$k_3 = p_5$, and $k_4 = -p_4$, where the external momenta satisfy the relation 
$p_1+p_2+p_3+p_4=p_5$. The diagram representing this integral is shown in 
Fig.~\ref{fig:feyn5p}.
\begin{figure}
\includegraphics[bb=140 560 310 730,clip=true,height=2.5in]{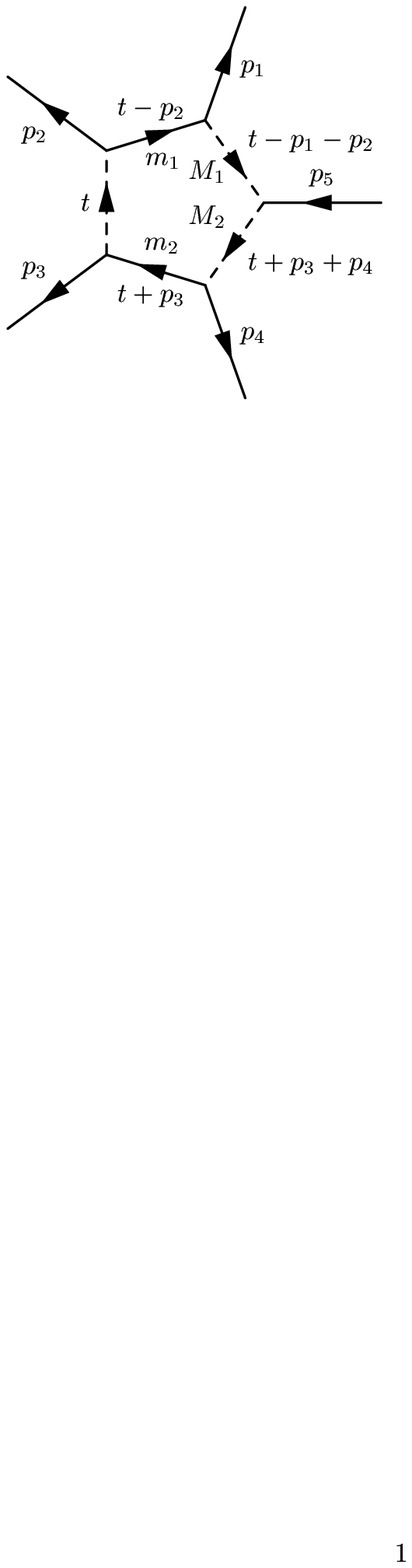}
\caption{\label{fig:feyn5p}Scalar 5--Point Function.}
\end{figure}

We have allowed the boson propagators to have arbitrary masses $M_1$ and
$M_2$ in order to include a form factor in the calculation of the 5--point
function. This is necessary since the form factor becomes a function of the
loop momentum. We will use a generalized DIP form factor which, with the 
appropriate choice of coefficients, can reproduce both the DIP and the BMS
form factor models. The generalized DIP form factor is
\begin{equation}
f(x_1,x_2) = 1 
 + \sum_i \alpha_i \left(  \frac{x_1}{x_1 - M_i^2/M^2} 
                         + \frac{x_2}{x_2 - M_i^2/M^2} \right)
 + \sum_{ij} \beta_{ij} \frac{x_1 x_2}
                             {(x_1 - M_i^2/M^2)(x_2 - M_j^2/M^2)},
\end{equation}
where $M$ is the total mass and the sum is over propagator masses $M_i$. 
The diagram containing two photon propagators and a form
factor is then replaced by a sum of $n^2$ diagrams of four different types, 
one containing two photon propagators, two containing one photon and one 
massive boson propagator, and one containing two massive boson propagators, 
\begin{equation}
\frac{f(x_1,x_2)}{x_1 x_2} = \frac{1}{x_1 x_2} 
 + \sum_i \alpha_i \left(  \frac{1}{x_1(x_2 - M_i^2/M^2)} 
                         + \frac{1}{x_2(x_1 - M_i^2/M^2)} \right)
 + \sum_{ij} \beta_{ij} \frac{1}{(x_1 - M_i^2/M^2)(x_2 - M_j^2/M^2)}.
\end{equation}
The general form of Eq.~(\ref{eq:i50}) permits the evaluation of all four 
of these contributions. To use a specific model, the parameters $\alpha_i$ and
$\beta_{ij}$ must be adjusted. For the simple DIP model, only a $\rho$ meson
term is included with $\alpha_1 = \alpha_{DIP}$ and $\beta_{11} =
\beta_{DIP}$. The DIP model requires the inclusion of four 5--point diagrams 
involving all combinations of photons and $\rho$ mesons. The BMS model is more
complicated, requiring 25 different diagrams. To simplify this we have
let $M_\omega = M_\rho$, which reduces the number of diagrams to 16. The
values of the generalized parameters, in terms of $\alpha_{K^*}$ are
\begin{subequations}
\begin{align}
\alpha_1 &= -1 + \frac{10 C \alpha_{K^*}}{9(1-M_\rho^2/M_{K^*}^2)},\\
\alpha_2 &= \frac{2 C \alpha_{K^*}}{9}
            \left(  \frac{1}{1-M_\phi^2/M_{K^*}^2}
                  - \frac{1}{1-M_\rho^2/M_{K^*}^2}\right),\\
\alpha_3 &= \frac{2 C \alpha_{K^*}}{1-M_\phi^2/M_{K^*}^2},
\end{align}
\end{subequations}
and $\beta_{ij} = \alpha_i \alpha_j$. A flat form factor is obtained by
setting all of the $\alpha_i = 0$ and $\beta_{ij} = 0$.

We will write the 5--point function in Eq.~(\ref{eq:i50}) as a sum of 
4--point functions using the following relationship between $n$--point 
functions and $(n-1)$--point functions~\cite{bern},
\begin{equation} \label{eq:sum1}
I_n = \frac{1}{2} \left[ 
   -\sum_{i=1}^n c_i I_{n-1}^{(i)} + (n-5+2\epsilon) c_0 I_n^{D=6-2\epsilon}
                  \right],
\end{equation}
where
\begin{align}
c_i    & = \sum_{j=1}^n \mathcal{S}_{ij}^{-1}, & 
c_0    & = \sum_{i=1}^n c_i = \sum_{i,j=1}^n \mathcal{S}_{ij}^{-1}, &
S_{ij} & = (\mu_i^2 + \mu_j^2 - k_{ij}^2)/2,\label{eq:smat}
\end{align}
and
\begin{align}
k_{ii} & = 0, &
k_{ij} & = k_i + k_{i+1} + \dots + k_{j-1}, &
\text{for}\quad & i < j.
\end{align}
We will use $\mu$ to refer to the propagator mass when the distinction
between vector bosons and leptons is irrelevant. 
In the case at hand 
$n=5$ so the second term in Eq.~(\ref{eq:sum1}) is $\mathcal{O}(\epsilon)$, 
and since the 5--point function in $D=6-2\epsilon$ dimensions is finite in 
the limit $\epsilon \to 0$, the scalar 5--point function can be written as 
a sum of five scalar 4--point functions
\begin{equation}
I_{50} = -\frac{1}{2} \sum_{i=1}^5 \left(\sum_{j=1}^5 S_{ij}^{-1}\right)
         I_{40}^{(i)} + \mathcal{O}(\epsilon).
\end{equation}

\paragraph{The 4--Point Function}%
Two of the five 4--point functions contain IR divergences due to the
presence of the 3--point functions where both of the vector boson propagators
have been removed. Therefore there are two distinct 4--point functions that
we will need. The first has one zero mass propagator 
\begin{multline} \label{eq:i40a}
I_{40}(k_1, k_2, k_3, \mu_1, \mu_2, \mu_3) = \\
  \mu^{2\epsilon} \int \frac{d^D t}{(2\pi)^D}
    \frac{1}{[t^2                       + i\epsilon]
             [(t+k_1)^2         - \mu_1^2 + i\epsilon]
             [(t+k_1+k_2)^2     - \mu_2^2 + i\epsilon]
             [(t+k_1+k_2+k_3)^2 - \mu_3^2 + i\epsilon]},
\end{multline}
and the other has two zero mass propagators and two lepton propagators
\begin{multline} \label{eq:i40b}
I_{40}^\prime(k_1, k_2, k_3, m_1, m_2) = \\
  \mu^{2\epsilon} \int \frac{d^D t}{(2\pi)^D}
    \frac{1}{[t^2                       + i\epsilon]
             [(t+k_1)^2                 + i\epsilon]
             [(t+k_1+k_2)^2     - m_1^2 + i\epsilon]
             [(t+k_1+k_2+k_3)^2 - m_2^2 + i\epsilon]},
\end{multline}
where in $I_{40}^\prime$, $(k_1+k_2)^2=m_1^2$ and $(k_1+k_2+k_3)^2=m_2^2$.

In order to use Eq.~(\ref{eq:i40a}) for $I_{40}^{(1)}$ where all four 
propagators have non--zero masses, and to extract the divergent part of
$I_{40}^{(3)}$ and $I_{40}^{(4)}$, we make use of the following propagator 
identity
\begin{equation} \label{eq:propagator}
\frac{1}{[(q+p)^2 - \mu_1^2][(q+p+k)^2 - \mu_2^2]} 
  = \frac{\alpha}{[(q+p)^2 - \mu_1^2](q+p+\alpha k)^2}
  + \frac{1-\alpha}{[(q+p+k)^2 - \mu_2^2](q+p+\alpha k)^2},
\end{equation}
where $\alpha$ is chosen to be the positive root of the equation
\begin{equation}
\alpha(1-\alpha) k^2 - (1-\alpha) \mu_1^2 - \alpha \mu_2^2 = 0.
\end{equation}
This identity allows us to write the five 4--point functions as
\begin{subequations}
\begin{align}
I_{40}^{(1)} & 
  = \alpha_1     I_{40}(-\alpha_1 p_{23}, -p_1, p_5, m_1, M_1, M_2)
  + (1-\alpha_1) I_{40}((1-\alpha_1) p_{23}, p_4-p_5, p_5, m_2, M_1, M_2),\\
I_{40}^{(2)} & = I_{40}(-p_{12}, p_5, -p_4, M_1, M_2, m_2),\\
I_{40}^{(3)} & 
  = \alpha_3     I_{40}^\prime(-\alpha_3 p_{34}, -p_2, p_{23}, m_1, m_2)
  + (1-\alpha_3) I_{40}((1-\alpha_3)p_{34}, p_1-p_5, p_{23}, M_2, m_1, m_2),\\
I_{40}^{(4)} & 
  = \alpha_4     I_{40}^\prime(\alpha_4 p_{12}, p_3, -p_{23}, m_2, m_1) 
  + (1-\alpha_4) I_{40}(-(1-\alpha_4) p_{12}, p_1, p_{23}, M_1, m_1, m_2),\\
I_{40}^{(5)} & = I_{40}(-p_2, -p_1, p_5, m_1, M_1, M_2),
\end{align}
\end{subequations}
where
\begin{align}
\alpha_1 & = (1 + \delta_{23} + \lambda_{23})/2, &
\alpha_3 & = 1 - M_2^2/p_{34}^2, &
\alpha_4 & = 1 - M_1^2/p_{12}^2.
\end{align}

The finite 4--point function $I_{40}$ defined in Eq.~(\ref{eq:i40a}) can be
expressed in closed form as a sum of 36 dilogarithms. We will define it
in terms of the function
\begin{align}
\mathcal{J}(A,B) & = \int_0^1 \frac{d z}{z - A} 
    \left[ \ln(z - B \pm i\epsilon) - \ln(A - B \pm i\epsilon) \right],\nn\\
  & = \dilog{\left(\frac{A}{A - B \pm i\epsilon}\right)}
    - \dilog{\left(\frac{A - 1}{A - B \pm i\epsilon}\right)}.
\end{align}
For arbitrary complex arguments, $A$ and $B$, the integration would also 
produce additional logarithms with prefactors which depend on the relative
difference between the signs of the imaginary parts of $A$ and 
$B$~\cite{thooft}, however 
if $A$ is real these additional terms vanish. In the case at hand, $A$ will
always be real and we will only need the dilogarithms. In terms of these 
new functions,
\begin{align}
I_{40}& (k_1, k_2, k_3, \mu_1, \mu_2, \mu_3) = 
 \frac{i}{16 \pi^2 \gamma (\eta^+ - \eta^-)}\nn\\& \times
  \bigg[
    \mathcal{J}\left(\beta+\eta^+, \frac{-e-d+i\epsilon}{k}\right)
  - \mathcal{J}\left(\beta+\eta^-, \frac{-e-d+i\epsilon}{k}\right)
  - \mathcal{J}\left(\frac{\eta^+}{1-\beta}, \frac{-d+i\epsilon}{e+k}\right)
\nn\\&
  + \mathcal{J}\left(\frac{\eta^-}{1-\beta}, \frac{-d+i\epsilon}{e+k}\right)
  + \mathcal{J}\left(\frac{-\eta^+}{\beta}, \frac{-d+i\epsilon}{e}\right)
  - \mathcal{J}\left(\frac{-\eta^-}{\beta}, \frac{-d+i\epsilon}{e}\right)
\nn\\&
  - \mathcal{J}\left(\beta+\eta^+, z_1^++i\epsilon\right)
  - \mathcal{J}\left(\beta+\eta^+, z_1^--i\epsilon\right)
  + \mathcal{J}\left(\beta+\eta^-, z_1^++i\epsilon\right)
  + \mathcal{J}\left(\beta+\eta^-, z_1^--i\epsilon\right)
\nn\\&
  + \mathcal{J}\left(\frac{\eta^+}{1-\beta}, z_2^++i\epsilon\right)
  + \mathcal{J}\left(\frac{\eta^+}{1-\beta}, z_2^--i\epsilon\right)
  - \mathcal{J}\left(\frac{\eta^-}{1-\beta}, z_2^++i\epsilon\right)
  - \mathcal{J}\left(\frac{\eta^-}{1-\beta}, z_2^--i\epsilon\right)
\nn\\&
  - \mathcal{J}\left(\frac{-\eta^+}{\beta}, z_3^++i\epsilon\right)
  - \mathcal{J}\left(\frac{-\eta^+}{\beta}, z_3^--i\epsilon\right)
  + \mathcal{J}\left(\frac{-\eta^-}{\beta}, z_3^++i\epsilon\right)
  + \mathcal{J}\left(\frac{-\eta^-}{\beta}, z_3^--i\epsilon\right)
  \bigg],
\end{align}
where $\beta$ is either root of the equation 
$g \beta^2 + j \beta + b = 0$, $\eta^\pm$ are the roots of the equation 
\begin{equation}
   [eg - jk - \beta gk] \eta^2
 + [eh - ck - dj - 2\beta dg] \eta
 + [ae - cd + \beta(ak - dh)] = 0,
\end{equation}
and $\gamma = eg - jk - \beta gk$. The quantities $z_i^\pm$ are the roots of
the following equations
\begin{subequations}
\begin{align}
0 & = g z_1^2 + (h + j + k) z_1 + (a + b + c + d + e),\\
0 & = (b + g + j) z_2^2 + (c + e + h + k) z_2 + (a + d),\\
0 & = b z_3^2 + (c + e) z_3 + (a + d).
\end{align}
\end{subequations}
The lower case variables are combinations of the elements of the relevant 
$4 \times 4$ matrix $S_{ij}$ defined in Eq.~(\ref{eq:smat}),
\begin{subequations}
\begin{align}
a & = S_{33} + S_{44} - 2 S_{34}           , &
f & = S_{44}                               , \\
b & = S_{22} + S_{33} - 2 S_{23}           , &
g & = S_{11} + S_{22} - 2 S_{12}           , \\
c & = 2 (S_{23} - S_{24} - S_{33} + S_{34}), &
h & = 2 (S_{13} - S_{14} - S_{23} + S_{24}),\\
d & = 2 (S_{34} - S_{44})                  , &
j & = 2 (S_{12} - S_{13} - S_{22} + S_{23}),\\
e & = 2 (S_{24} - S_{34})                  , &
k & = 2 (S_{14} - S_{24})                  .
\end{align}
\end{subequations}

The divergent 4--point function $I_{40}^\prime$ defined in
Eq.~(\ref{eq:i40b}) can also be written in closed from. The divergent part
is just the divergent 3--point function
\begin{align} \label{eq:iir}
I_{IR} & = \int \frac{d^4 t}{(2\pi)^4}
 \frac{1}{[t^2   - \Lambda^2 + i\epsilon]
          [(t-p_2)^2 - m_1^2 + i\epsilon]
          [(t+p_3)^2 - m_2^2 + i\epsilon]},\nn\\
 & = \frac{i}{32 \pi^2 p_{23}^2 \lambda_{23}} 
     \bigg\{
         \bigg[
          \ln{\left(\frac{z_{23}+\lambda_{23}}{z_{23}-\lambda_{23}}\right)}
          - 2 i\pi
         \bigg] \ln\frac{p_{23}^2}{\Lambda^2}
     - 2 \pi^2
     + 2 \dilog{\left(\frac{2 \lambda_{23}}{1+\delta_{23}+\lambda_{23}}\right)}
     + 2 \dilog{\left(\frac{2 \lambda_{23}}{1-\delta_{23}+\lambda_{23}}\right)}\nn\\
 & \quad \quad \quad \quad
     - 4 i \pi \ln \lambda_{23}
     - \ln{\left(\frac{1+\delta_{23}-\lambda_{23}}{2 \lambda_{23}}\right)}
       \ln{\left(\frac{1+\delta_{23}+\lambda_{23}}{2}\right)}
     - \ln{\left(\frac{1-\delta_{23}-\lambda_{23}}{2 \lambda_{23}}\right)}
       \ln{\left(\frac{1-\delta_{23}+\lambda_{23}}{2}\right)}\nn\\
 & \quad \quad \quad \quad
     + \ln^2{\left(\frac{1-\delta_{23}+\lambda_{23}}{2 \lambda_{23}}\right)} 
     + \ln^2{\left(\frac{1+\delta_{23}+\lambda_{23}}{2 \lambda_{23}}\right)} 
     + \ln{\lambda_{23}} \ln{\left(\frac{1-\delta_{23}+\lambda_{23}}
                                        {2 \lambda_{23}}\right)} 
     + \ln{\lambda_{23}} \ln{\left(\frac{1+\delta_{23}+\lambda_{23}}
                                        {2 \lambda_{23}}\right)}\nn\\
 & \quad \quad \quad \quad
     + \frac{1}{2} \bigg[ 
                      \ln^2{\left(\frac{1+\delta_{23}+\lambda_{23}}{2}\right)}
                    + \ln^2{\left(\frac{1-\delta_{23}+\lambda_{23}}{2}\right)}
                    - \ln^2{\left(\frac{1+\delta_{23}-\lambda_{23}}{2}\right)}
                    - \ln^2{\left(\frac{1-\delta_{23}-\lambda_{23}}{2}\right)}
                   \bigg]
     \bigg\},
\end{align}
The full expression is then
\begin{align}
I_{40}^\prime(k_1, k_2, k_3, m_1, m_2) & = 
\frac{1}{k_1^2} \bigg(
  I_{IR} + \frac{i}{16 \pi^2 \lambda_{23} k_3^2} \bigg\{
   \ln{\left(\frac{2 k_1 \cdot k_3}{k_1^2}\right)}
    \left[\ln{\left(\frac{z_{23} - \lambda_{23}}{z_{23} + \lambda_{23}}\right)}
          + 2 i \pi
    \right]
 \nn\\& \quad \quad
 + \dilog{\left(\frac{1-A}{B-A+i\epsilon}\right)}
 - \dilog{\left(\frac{1-A}{C-A-i\epsilon}\right)}
 - \dilog{\left(\frac{-A}{B-A+i\epsilon}\right)}
 + \dilog{\left(\frac{-A}{C-A-i\epsilon}\right)}
 \nn\\& \quad \quad
 + \ln(1-A-i\epsilon) \left[  \ln(1-B-i\epsilon)
                            - \ln(1-C+i\epsilon)
                            - \ln(A-B-i\epsilon)
                            + \ln(A-C+i\epsilon)\right]
 \nn\\& \quad \quad
 - \ln(-A-i\epsilon)  \left[  \ln(-B-i\epsilon)
                            - \ln(-C+i\epsilon)
                            - \ln(A-B-i\epsilon)
                            + \ln(A-C+i\epsilon)\right]
\bigg\} \bigg),\label{eq:i40p}
\end{align}
where 
\begin{align}
A & = -(k_1^2/2 + k_1 \cdot k_2)/(k_1 \cdot k_3), \label{eq:aa}&
B & = (1+\delta_{23}+\lambda_{23})/2, &
C & = (1+\delta_{23}-\lambda_{23})/2.
\end{align}

It is worthwhile to consider the special case of two photon propagators,
that is $M_1 = M_2 = 0$. In that case the 4--point functions simplify
considerably. We will write
\begin{align}
I_{40}^{\dagger}(k_1,k_2,k_3,\mu_1,\mu_2) &= \int \frac{d^4 t}{(2\pi)^4}
  \frac{1}{[t^2 - \mu_1^2 + i\epsilon]
           [(t+k_1)^2 - \mu_2^2 + i\epsilon]
           [(t+k_1+k_2)^2 + i\epsilon]
           [(t+k_1+k_2+k_3)^2 + i\epsilon]},\nn\\
 &= \frac{i}{\sqrt{\Delta}} \bigg\{ 
   \bigg[
    \dilog{\left(\frac{v^+}    {v^+ - r_1}                     \right)}
  - \dilog{\left(\frac{v^+ - 1}{v^+ - r_1}                     \right)}
  - \dilog{\left(\frac{v^+ - 1}{v^+ - r_2}                     \right)}
  - \dilog{\left(\frac{v^+}    {v^+ - \tilde{v}^+ - i \epsilon}\right)}
 \nn\\ & \quad\quad\quad\quad
  + \dilog{\left(\frac{v^+ - 1}{v^+ - \tilde{v}^+ - i \epsilon}\right)} 
  - \dilog{\left(\frac{v^+}    {v^+ - \tilde{v}^- + i \epsilon}\right)}
  + \dilog{\left(\frac{v^+ - 1}{v^+ - \tilde{v}^- + i \epsilon}\right)}
   \bigg]\nn\\
 & \quad\quad\quad
 - \bigg[
    \dilog{\left(\frac{v^-}    {v^- - r_1}                     \right)}
  - \dilog{\left(\frac{v^- - 1}{v^- - r_1}                     \right)}
  - \dilog{\left(\frac{v^- - 1}{v^- - r_2}                     \right)}
  - \dilog{\left(\frac{v^-}    {v^- - \tilde{v}^+ - i \epsilon}\right)}
 \nn\\ & \quad\quad\quad\quad
  + \dilog{\left(\frac{v^- - 1}{v^- - \tilde{v}^+ - i \epsilon}\right)} 
  - \dilog{\left(\frac{v^-}    {v^- - \tilde{v}^- + i \epsilon}\right)}
  + \dilog{\left(\frac{v^- - 1}{v^- - \tilde{v}^- + i \epsilon}\right)}
   \bigg]\bigg\},
\end{align}
where $\Delta$ is the discriminant and $v^\pm$ are the roots of the
quadratic equation
\begin{equation}
[k (h+k) - dg] v^2 + [e (h+k) - dj] v - bd = 0,
\end{equation}
and $\tilde{v}^\pm$ are the roots of
\begin{equation}
g \tilde{v}^2 + (h+j+k) \tilde{v} + b = 0,
\end{equation}
and
\begin{align}
r_1 &= (-d-e+i\epsilon)/k, & r_2 &= i\epsilon/(h+k).
\end{align}
Therefore, when $M_1 = M_2 = 0$, the five 4--point functions are simply
\begin{subequations}
\begin{align}
I_{40}^{(1)} &= I_{40}^\dagger(p_{23}, p_4, -p_5, m_1, m_2),\\
I_{40}^{(2)} &= I_{40}^\dagger(p_3, p_4, -p_5, 0, m_2),\\
I_{40}^{(3)} &= I_{40}^\prime(-p_{34}, -p_2, p_{23}, m_1, m_2),\\
I_{40}^{(4)} &= I_{40}^\prime(p_{12}, p_3, -p_{23}, m_2, m_1),\\
I_{40}^{(5)} &= I_{40}^\dagger(-p_2, -p_1, p_5, 0, m_1).
\end{align}
\end{subequations}
Also, in this case, $I_{40}^\prime$ simplifies somewhat because $A$ defined
in Eq.~(\ref{eq:aa}) becomes one. The first two dilogarithms in the second
line of Eq.~(\ref{eq:i40p}), along with the entire third line, vanish in
this case. $I_{40}^\prime$ then becomes
\begin{align}
I_{40}^\prime(k_1, k_2, k_3, m_1, m_2) & = 
\frac{1}{k_1^2} \bigg(
  I_{IR} + \frac{i}{16 \pi^2 \lambda_{23} k_3^2} \bigg\{
   \ln{\left(\frac{2 k_1 \cdot k_3}{k_1^2}\right)}
    \left[\ln{\left(\frac{z_{23} - \lambda_{23}}{z_{23} + \lambda_{23}}\right)}
          + 2 i \pi
    \right]
 \nn\\& \quad \quad
 - \dilog{\left(\frac{-1}{B-1+i\epsilon}\right)}
 + \dilog{\left(\frac{-1}{C-1-i\epsilon}\right)}
 \nn\\& \quad \quad
 +i \pi  \left[  \ln(-B-i\epsilon)  - \ln(-C+i\epsilon)
               - \ln(1-B-i\epsilon) + \ln(1-C+i\epsilon)\right]
\bigg\} \bigg),
\end{align}
where $B$ and $C$ are still given by Eq.~(\ref{eq:aa}).

\paragraph{The 3--Point Function}%
There are ten 3--point functions that are needed, all of which are finite
except one, $I_{30}^{(34)} = I_{IR}$ defined in Eq.~(\ref{eq:iir}).
The superscripts used in this section denote the two propagators that have
been dropped from the original 5--point function to obtain the particular
3--point function. The finite 3--point functions can be generically 
written as
\begin{align}
I_{30}(k_1, k_2, \mu_1, \mu_2, \mu_3) 
 & = \int \frac{d^D t}{(2\pi)^D} 
          \frac{1}{[t^2           - \mu_1^2 + i\epsilon]
                   [(t+k_1)^2     - \mu_2^2 + i\epsilon]
                   [(t+k_1+k_2)^2 - \mu_3^3 + i\epsilon]},\\
 & = \frac{-i}{16 \pi^2 (c + 2 b \beta)} \bigg\{ 
   \bigg[
     \dilog{\left(\frac{v_1}    {v_1 - \tilde{v}_1^+ - i \epsilon}\right)}
   - \dilog{\left(\frac{v_1 - 1}{v_1 - \tilde{v}_1^+ - i \epsilon}\right)}
   + \dilog{\left(\frac{v_1}    {v_1 - \tilde{v}_1^- + i \epsilon}\right)}
\nn\\ & \quad \quad \quad \quad \quad \quad \quad
   - \dilog{\left(\frac{v_1 - 1}{v_1 - \tilde{v}_1^- + i \epsilon}\right)}
   - \dilog{\left(\frac{v_2}    {v_2 - \tilde{v}_2^+ - i \epsilon}\right)}
   + \dilog{\left(\frac{v_2 - 1}{v_2 - \tilde{v}_2^+ - i \epsilon}\right)}
\nn\\ & \quad \quad \quad \quad \quad \quad \quad
   - \dilog{\left(\frac{v_2}    {v_2 - \tilde{v}_2^- + i \epsilon}\right)}
   + \dilog{\left(\frac{v_2 - 1}{v_2 - \tilde{v}_2^- + i \epsilon}\right)}
   + \dilog{\left(\frac{v_3}    {v_3 - \tilde{v}_3^+ - i \epsilon}\right)}
\nn\\ & \quad \quad \quad \quad \quad \quad \quad
   - \dilog{\left(\frac{v_3 - 1}{v_3 - \tilde{v}_3^+ - i \epsilon}\right)}
   + \dilog{\left(\frac{v_3}    {v_3 - \tilde{v}_3^- + i \epsilon}\right)}
   - \dilog{\left(\frac{v_3 - 1}{v_3 - \tilde{v}_3^- + i \epsilon}\right)}
   \bigg] \bigg\},
\end{align}
where the $v_i$ are
\begin{align}
v_1 & = -\frac{2a + d + \beta(c + e)}{c + 2 b \beta},  &
v_2 & = -\frac{d + e \beta}{(1-\beta)(c + 2 b \beta)}, &
v_3 & =  \frac{d + e \beta}{\beta(c + 2 b \beta)},
\end{align}
and the $\tilde{v}_i^\pm$ are roots of the three quadratic equations
\begin{subequations}
\begin{align}
0 & = b \tilde{v}_1^2 + (c + e) \tilde{v}_1 + (a + d + f),\\
0 & = (a + b + c) \tilde{v}_2^2 + (d + e) \tilde{v}_2 + f,\\
0 & = a \tilde{v}_3^2 + d \tilde{v}_3 + f,
\end{align}
\end{subequations}
and $\beta$ is either root of the equation 
$b \beta^2 + c \beta + a = 0$.
The lower case letters are again combination of the relevant matrix $S_{ij}$
\begin{subequations}
\begin{align}
a & = S_{22} + S_{33} - 2 S_{23}           , &
c & = 2 (S_{12} - S_{13} - S_{22} + S_{23}), &
e & = 2 (S_{13} - S_{23})                  , \\
b & = S_{11} + S_{22} - 2 S_{12}           , &
d & = 2 (S_{23} - S_{33})                  , &
f & = S_{33}.
\end{align}
\end{subequations}

The ten 3--point functions that we need can then be expressed as
\begin{subequations}
\begin{align}
I_{30}^{(12)} & = I_{30}(    p_4,    -p_5, m_2, M_2, M_1), &
I_{30}^{(24)} & = I_{30}(    p_3,-p_5+p_4,   0, m_2, M_1), \\
I_{30}^{(13)} & = I_{30}( p_{23},     p_4, m_1, m_2, M_2), &
I_{30}^{(25)} & = I_{30}(-p_{12},     p_5,   0, M_1, M_2), \\
I_{30}^{(14)} & = I_{30}( p_{23},-p_5+p_4, m_1, m_2, M_1), &
I_{30}^{(34)} & = I_{IR}                                 , \\
I_{30}^{(15)} & = I_{30}(   -p_1,     p_5, m_1, M_1, M_2), &
I_{30}^{(35)} & = I_{30}(   -p_2, p_5-p_1,   0, m_1, M_2), \\
I_{30}^{(23)} & = I_{30}(    p_3,     p_4,   0, m_2, M_2), &
I_{30}^{(45)} & = I_{30}(   -p_2,    -p_1,   0, m_1, M_1).
\end{align}
\end{subequations}

\paragraph{The 2--Point Function}%
And finally, the general expression for the 2--point function is
\begin{align}
I_{20}(k_1, \mu_1, \mu_2)
 & = \int \frac{d^D t}{(2\pi)^D}
          \frac{1}{[t^2       - \mu_1^2 + i\epsilon]
                   [(t+k_1)^2 - \mu_2^2 + i\epsilon]},\\
 & = \frac{i}{(4 \pi)^2} \bigg[
      \Gamma(\varepsilon) \left(\frac{4 \pi \mu^2}{k_1^2}\right)^\varepsilon
    + 2 - (1-v^+)\ln{(1-v^+-i\epsilon)} - (1-v^-)\ln{(1-v^-+i\epsilon)}\nn\\
 & \quad \quad \quad \quad \quad
    - v^+\ln{(-v^+ -i\epsilon)} - v^-\ln{(-v^- +i\epsilon)} \bigg],
\end{align}
where $v^\pm$are roots to the quadratic equation
\begin{equation}
k_1^2 v^2 + (\mu_1^2 - \mu_2^2 - k_1^2) v + \mu_2^2 = 0.
\end{equation}
The UV divergent term containing $\varepsilon$ cancels when the 2--point 
functions are combined to form the tensor integrals and can therefore be 
safely ignored. The ten 2--point functions that we require are then
\begin{subequations}
\begin{align}
I_{20}^{(123)} & = I_{20}(    -p_4, M_2, m_2), &
I_{20}^{(145)} & = I_{20}(    -p_1, m_1, M_1), \\
I_{20}^{(124)} & = I_{20}( p_5-p_4, M_1, m_2), &
I_{20}^{(234)} & = I_{20}(     p_3,   0, m_2), \\
I_{20}^{(125)} & = I_{20}(     p_5, M_1, M_2), &
I_{20}^{(235)} & = I_{20}(  p_{34},   0, M_2), \\
I_{20}^{(134)} & = I_{20}(  p_{23}, m_1, m_2), &
I_{20}^{(245)} & = I_{20}( -p_{12},   0, M_1), \\
I_{20}^{(135)} & = I_{20}( p_5-p_1, m_1, M_2), &
I_{20}^{(345)} & = I_{20}(    -p_2,   0, m_1).
\end{align}
\end{subequations}

\end{widetext}

\bibliography{prd_dbldal}

\end{document}